\documentclass[a4paper,11pt]{article}
\usepackage{jheppub,mathrsfs,amsmath,amssymb,bm} 
\usepackage{lineno}
\usepackage{subfig}
\usepackage{braket}

\title{\boldmath More on OTOCs and Chaos in Quantum Mechanics - Magnetic Fields}

\author{Cameron Beetar \& Jeff Murugan}
\affiliation{The Laboratory for Quantum Gravity \& Strings,\\
Department of Mathematics and Applied Mathematics,\\
University of Cape Town, Private Bag, Rondebosch, 7701,\\
South Africa}

\emailAdd{jeff.murugan@uct.ac.za}

\abstract{We revisit thermal out-of-time-order correlators (OTOCs) in single-particle quantum systems, focusing on magnetic billiards. Using the stadium billiard as a testbed, we compute the thermal OTOC $C_T(t) = -\langle [x(t), p]^2 \rangle_\beta$ and extract Lyapunov-like exponents $\lambda_L$ that quantify early-time growth. We map out $\lambda_L(T, B)$, revealing a crossover from quantum chaos to magnetic rigidity. In parallel, we compute an alternative OTOC built from guiding-center operators, which exhibits qualitatively distinct dynamics and no exponential growth. Our results offer a controlled framework for probing scrambling, temperature dependence, and the interplay of geometry and magnetic fields in quantum systems.}

\begin{document}
\maketitle
\flushbottom

\section{Introduction}
\label{sec:intro}
Few things shape the quantum realm as profoundly as magnetic fields, influencing everything from the Landau quantization of electron orbits to the emergence of topological matter and quantum Hall effects. When combined with Out-of-Time-Ordered Correlators (OTOCs) which have already proven to be powerful probes of quantum information scrambling and chaos, the interplay between quantum magnetism and information dynamics presents a fertile ground for new insights.\\

\noindent
A key tool in this exploration, at least in the context of single-particle quantum systems, is the OTOC construction introduced \cite{Hashimoto:2017oit}, which has provided a useful framework to diagnose quantum chaos in isolated single-particle quantum systems. The formalism there builds on the idea that in classically chaotic systems, small perturbations grow exponentially, a signature usually captured by the Lyapunov exponent in classical mechanics. While the idea that the OTOC,

\begin{eqnarray}
C_T(t) = \frac{1}{Z} \sum_n e^{-\beta E_n} \langle n | [x(t), p]^2 | n \rangle,
\label{OTOC}
\end{eqnarray}

\noindent
serves as a quantum analog of this behavior and exhibits exponential growth in quantum chaotic systems is not new, the work of Hashimoto {\it et.al.} showed how, given the spectrum of a single-particle Hamiltonian and the matrix elements of a class of simple operators, the OTOC in \eqref{OTOC} can be constructed by summing over a set of microcanonical OTOCs\footnote{When discussing thermal OTOCs derived from the summation of microcanonical OTOCs, we will use `thermal' and `canonical' interchangeably.}. This in turn was used to study a number of simple single-particle systems with well-known spectra, such as the 1-dimensional particle in a box and harmonic oscillator. The construction was also extended to higher dimensions and non-integrable Hamiltonians; in particular systems, like 2-dimensional stadium billiards that are classically chaotic.\\ 

\noindent
While OTOCs have been widely used to diagnose quantum chaos in many-body systems, where they exhibit exponential growth similar to classical Lyapunov behavior, an intriguing result from Hashimoto et al.~\cite{Hashimoto:2017oit} is that single-particle quantum systems, even those with classically chaotic counterparts, do not show this behavior. Instead, OTOCs in such systems typically exhibit power-law growth at early times, followed by saturation at long times, fundamentally different from the unbounded growth seen in many-body systems. This distinction arises because, while classical chaos is characterized by exponential sensitivity to initial conditions, quantum mechanics replaces trajectory divergence with unitary wavefunction evolution, suppressing classical-like exponential separation. Moreover, the finite-dimensional Hilbert space of single-particle quantum systems prevents indefinite operator growth, leading to OTOC saturation. \\

\noindent
Beyond these broad features, OTOCs in single-particle systems reveal a range of dynamical behaviors that provide valuable insights into quantum chaos, spectral statistics, and quantum transport. In integrable systems, such as the hydrogen atom, OTOCs can exhibit oscillatory recurrences, reflecting the periodic motion of wavefunctions in phase space~\cite{Berry:1977pr}. By contrast, in classically chaotic systems such as quantum billiards, OTOCs display power-law growth, signaling quantum wavefunction delocalization rather than trajectory divergence~\cite{Gutzwiller:1990ch}. The crucial timescale that dictates this transition is the {\it Ehrenfest time} \( t_E \), the point at which classical evolution gives way to quantum interference effects. For times \( t < t_E \), OTOCs may follow semiclassical growth patterns, but for \( t > t_E \), quantum interference dominates, washing out classical-like chaos~\cite{Zaslavsky:1981eh,Shepelyansky:1994hy}.\\

\noindent
Interestingly, while OTOCs do not directly exhibit chaos in the form of exponential growth in these systems, they do correlate with spectral chaos, providing a dynamical probe of Wigner-Dyson level statistics~\cite{Haake:2001qp,Mehta:2004sp}. In chaotic billiards \cite{CASATI1999293}, for instance, the late-time saturation of OTOCs may be linked to universality classes of quantum spectra. Similarly, in confined chaotic geometries, OTOCs provide a direct window into wavefunction scarring, where quantum states localize along unstable classical periodic orbits~\cite{Heller:1984ws}. More broadly, OTOCs also serve as a tool to study quantum transport, particularly in settings where classical chaos plays a role in diffusive behavior~\cite{Chirikov:1981qm}.\\

\noindent
An important yet underexplored frontier for OTOCs lies in quantum systems subject to transverse magnetic fields. Such fields fundamentally reshape the classical phase space by introducing cyclotron motion, modifying ergodicity and, in the strong-field regime, driving Landau quantization. This reorganization of energy levels and spatial structure localizes wavefunctions within tight cyclotron orbits, effectively suppressing spatial overlap and long-range coherence. These features raise some natural: How does a magnetic field affect quantum information scrambling? Does it suppress, enhance, or qualitatively reshape the growth of OTOCs? In this work, we systematically extend the construction of Hashimoto et al.~\cite{Hashimoto:2017oit} to magnetized quantum systems, using OTOCs to probe the interplay between quantum chaos, Landau physics, and thermal scrambling. Our analysis is guided by three key questions:
\begin{enumerate}
    \item How are quantum chaos and scrambling modified by magnetic fields? Even though free-particle motion in a magnetic field is integrable, confined systems—such as magnetized billiards—can still exhibit rich chaotic dynamics. The OTOC provides a quantitative lens to assess whether magnetic confinement suppresses or reorganizes chaotic behavior.
	\item What role does Landau quantization play in information flow? As the magnetic field strengthens, Landau levels emerge and classical trajectories become tightly coiled, localizing wavefunctions and reducing phase space transport. This suggests a potential slowing down of scrambling and a transition toward more rigid, non-chaotic dynamics. Moreover, the resulting noncommutative geometry of guiding centers may imprint new structures in the OTOC.
	\item How do thermal and microcanonical OTOCs behave in this context? At finite temperature, the magnetic field reshapes the thermal ensemble by altering level spacing and occupation. We investigate whether and how this affects the early-time growth and long-time saturation of thermal OTOCs, and compare it to the behavior of microcanonical versions constructed from Landau-projected states.
\end{enumerate}
We focus on thermal OTOCs of both canonical operators (position and momentum) and guiding-center observables, examining their growth behavior in a stadium billiard geometry subjected to a transverse magnetic field. Our approach combines exact diagonalization of the magnetic Hamiltonian, construction of thermal correlators, and early-time fitting of OTOC profiles to extract Lyapunov-like exponents. In particular, we chart the joint dependence of scrambling on temperature and magnetic field strength, and construct a surface $\lambda_L(T, B)$ that characterizes the transition from chaotic to magnetically rigid dynamics.\\

\noindent
The remainder of the paper is organized as follows: Section 2 reviews the setup and the construction of OTOCs in quantum mechanics with illustrative examples; Section 3 presents our numerical methods and results for canonical OTOCs in magnetised quantum systems; Section 4 introduces the guiding-center formulation and compares the two diagnostics; and Section 5 concludes with a discussion of broader implications and future directions.	

\section{OTOCs in quantum mechanics}
\label{sec:qm_otoc}
Let's start by reviewing the formalism developed in \cite{Hashimoto:2017oit} to compute OTOCs for a general time-independent Hamiltonian  $H = H(\bm{x}_{i},\bm{p}_{i})$. If the system under consideration is at temperature $T = 1/\beta$, then we define the {\it thermal} out-of-time-order correlator as
\begin{eqnarray}\label{thermal otoc}
    C_{T}(t) = -\langle [x(t),p(0)]^{2}\rangle\,,
\end{eqnarray}
where the thermal expectation value of the operator $O$ is $\langle O\rangle
\equiv \mathrm{Tr}\left(e^{-\beta H}O\right)/\mathrm{Tr}\,e^{-\beta H}$ and $x(t) = e^{iHt}x(0)\,e^{-iHt}$ is the Heisenberg operator associated to the Schr\"odinger operator\footnote{Following the example of \cite{Hashimoto:2017oit}, we will also drop the argument of Heisenberg operators at $t=0$ so, for example, $x\equiv x(0)$ {\it etc.}} $x(0)$. In the energy eigenstate basis for the Hilbert space, 
\eqref{thermal otoc} can be expanded as
\begin{eqnarray}
    C_{T} = -\frac{1}{Z}\sum_{n}e^{-\beta E_{n}}\langle n|[x(t),p]^{2}|n\rangle \equiv \frac{1}{Z}\sum_{n}e^{-\beta E_{n}}c_{n}\,,
\end{eqnarray}
where, as usual, $H|n\rangle = E_{n}|n\rangle$ and $c_{n}$, the OTOC in a fixed energy eigenstate, is referred to as a {\it microcanonical} OTOC. The key feature of the Hashimoto {\it et.al.} construction is that, given the matrix elements of the position operator and the spectrum of the theory, the microcanonical, and by re-summation, the thermal OTOCs can be computed, if not exactly, then at least numerically. To see how, note that the microcanonical OTOC
\begin{eqnarray}
    c_{n}(t) = - \langle n|[x(t),p]^{2}|n\rangle = \sum_{m}b_{nm}b_{nm}^{*}\,,
\end{eqnarray}
where we have used the fact that the energy eigenbasis is complete ({\it i.e.} $\sum_{m}|m\rangle\langle m| = \bm{1}$) and defined the Hermitian matrix element $b^{*}_{mn} = b_{nm}\equiv -i\langle n| [x(t),p]|m\rangle$. Expanding the commutator, defining the energy differences $E_{nm} \equiv E_{n} - E_{m}$, and inserting another resolution of the identity, we can write
\begin{eqnarray}
    b_{nm} = -i\sum_{k}\left(e^{iE_{nk}t}x_{nk}p_{km} - e^{iE_{km}t}p_{nk}x_{km}\right)\,,
\end{eqnarray}
where $x_{nm}$ and $p_{km}$ are matrix elements of the Schr\"odinger picture position and momentum operators respectively and, as usual, the Heisenberg and Schr\"odinger operators are related through $x(t) = e^{iHt}xe^{-iHt}$. While convenient, this expression is not particularly suitable for numerical computation since it involves explicit derivatives of wave functions. Fortunately, for a Hamiltonian of the standard form, $H = \bm{p}_{i}\cdot \bm{p}_{i} + U(\bm{x}_{i})$, the commutator $[H,x] = -2ip$ means that we can express the matrix elements of the momentum operator in terms of the position operator as, $p_{mn} = \frac{i}{2}E_{mn}x_{mn}$ so that
\begin{eqnarray}
    b_{nm} = \frac{1}{2}\sum_{k}x_{nk}x_{km}\left(E_{km}e^{iE_{nk}t} - E_{nk}e^{iE_{km}t}\right)\,.
    \label{eq:bnm}
\end{eqnarray}
\noindent
\subsection{A 2-dimensional harmonic oscillator}

To test the microcanonical OTOC construction introduced in~\cite{Hashimoto:2017oit}, we will first consider the exactly solvable two-dimensional quantum harmonic oscillator. As simple as this system is, it serves as an ideal integrable benchmark in which to validate spectral OTOC representations and explore their physical content in the absence of quantum chaos.\\

\noindent
The Hamiltonian is given by
\begin{equation}
    H = \frac{p_x^2 + p_y^2}{2m} + \frac{1}{2} m \omega^2 (x^2 + y^2),
\end{equation}
with eigenstates $|n_x, n_y\rangle$ and energy eigenvalues
\begin{equation}
    E_{n_x, n_y} = \hbar \omega (n_x + n_y + 1)\,, \quad n_x, n_y \in \mathbb{N}_0\,.
\end{equation}
We evaluate the microcanonical OTOC for the operator pair $(x(t), p_x)$ using both the Heisenberg approach and the spectral representation.\\

\noindent
In the Heisenberg picture, the time evolution of the $x$-coordinate is
\begin{equation}
    x(t) = x(0) \cos(\omega t) + \frac{p_x(0)}{m \omega} \sin(\omega t)\,,
\end{equation}
which implies that the commutator
\begin{equation}
    [x(t), p_x] = i \hbar \cos(\omega t), \quad \Rightarrow \quad [x(t), p_x]^2 = -\hbar^2 \cos^2(\omega t)\,.
\end{equation}
Therefore, the microcanonical OTOC for any eigenstate $|n\rangle$ is readily evaluated as
\begin{equation}
    c_n(t) = \hbar^2 \cos^2(\omega t),
\end{equation}
showing the clean, periodic behavior characteristic of an integrable system. It will also prove instructive to evaluate the microcanonical OTOC via
\begin{equation}
    c_n(t) = \sum_m |b_{nm}(t)|^2,
\end{equation}
with
\begin{equation}\label{Bmn}
    b_{nm}(t) = \frac{1}{2} \sum_k x_{nk} x_{km} \left( E_{km} e^{i E_{nk} t/\hbar} - E_{nk} e^{i E_{km} t/\hbar} \right),
\end{equation}
where $x_{nm} = \langle n | x | m \rangle$. For the harmonic oscillator, the position operator $x$ has tridiagonal matrix elements,
\begin{equation}
    \langle n_x \pm 1 | x | n_x \rangle = \sqrt{\frac{\hbar}{2m\omega}} \sqrt{n_x + 1}, \quad \text{or} \quad \sqrt{n_x}.
\end{equation}
Restricting to the $x$ sector and considering the ground state $|n_x=0\rangle$, we find that the only non-vanishing contribution to $c_0(t)$ comes from intermediate states connected by two applications of $x$, i.e., $|0\rangle \to |1\rangle \to |2\rangle$. This yields
\begin{equation}
    c_0(t) = \frac{\hbar^4}{2 m^2 \omega^2} \cos^2(\omega t),
\end{equation}
in agreement with the functional form from the Heisenberg method, but differing by a prefactor. The apparent mismatch in prefactors arises from differing conventions: the spectral method computes the OTOC using raw matrix elements of $x$, which carry dimensions of length. These enter quadratically and combine with energy differences, leading to the observed mass- and frequency-dependent prefactor. In contrast, the direct Heisenberg method uses the canonical commutation relations, whose structure is independent of such scales.\\

\noindent
To reconcile the two, we may define a dimensionless operator
\begin{equation}
    \tilde{x} = \sqrt{\frac{m \omega}{\hbar}} x,
\end{equation}
in terms of which the commutator becomes
\begin{equation}
    [\tilde{x}(t), p_x] = i \sqrt{m \omega \hbar} \cos(\omega t),
\end{equation}
and the corresponding OTOC recovers the normalization of the spectral representation. Alternatively, interpreting the spectral result as the OTOC of the unnormalized operator $x$ naturally acquires the dimensional prefactor.\\

\noindent
As defined above, the thermal OTOC is computed as the canonical ensemble average over the microcanonical OTOCs,
\begin{equation}
    C_T(t) = \frac{1}{Z} \sum_n e^{-\beta E_n} c_n(t), \quad Z = \sum_n e^{-\beta E_n},
\end{equation}
where $E_n = \hbar \omega (n_x + n_y + 1)$ and $c_n(t)$ is the microcanonical OTOC computed above. Using the fact that $c_n(t) = \hbar^2 \cos^2(\omega t)$ for all eigenstates (from the Heisenberg approach), the thermal OTOC becomes
\begin{equation}
    C_T(t) = \hbar^2 \cos^2(\omega t),
\end{equation}
since the $\beta$-weighted average of a constant yields the same constant. This is a peculiarity of the harmonic oscillator spectrum; the OTOC is independent of the quantum numbers, and consequently, the thermal average is trivial. However, when working via the spectral representation, each $c_n(t)$ has a mass- and frequency-dependent prefactor:
\begin{equation}
    c_n(t) = \frac{\hbar^4}{2 m^2 \omega^2} \cos^2(\omega t),
\end{equation}
so the thermal average becomes:
\begin{equation}
    C_T(t) = \frac{\hbar^4}{2 m^2 \omega^2} \cos^2(\omega t),
\end{equation}
again independent of temperature because all $c_n(t)$ are identical.\\

\noindent
This behavior is special to the harmonic oscillator due to its evenly spaced spectrum and the uniformity of the operator matrix elements across states. In more general systems, where $c_n(t)$ varies significantly with $n$, the thermal OTOC will exhibit nontrivial temperature dependence. In particular, in systems with more complex dynamics (e.g., those exhibiting chaos), the thermal OTOC is expected to show signatures of exponential growth at early times and eventual saturation—features entirely absent in this integrable model.\\

\noindent
This comparison reinforces the utility of the thermal OTOC in smoothing over state-specific fluctuations in $c_n(t)$ and offers an observable that is directly sensitive to the thermal population of levels. In chaotic systems, this averaging captures the interplay between energy-dependent scrambling rates and the thermally accessible portion of the spectrum.\\

\subsection{The Morse Potential}
\noindent
As a second illustration of the spectral approach to out-of-time-order correlators (OTOCs), we compute both microcanonical and thermal OTOCs for a quantum particle in the {\it Morse potential}. Less trivial than the harmonic oscillator, the Morse potential,
\begin{eqnarray}
    V(x) = D_e \left(1 - e^{-a(x - x_0)}\right)^2\,,
\end{eqnarray}
where $D_e$ is the depth of the potential, $a$ sets the confinement scale, and $x_0$ is the equilibrium position. It is a well-known integrable system that captures essential features of molecular vibrations.  It represents an anharmonic oscillator that supports a finite number of bound states, making it an ideal model for molecular vibrational modes.\\

\noindent
The bound-state energy levels of the Morse potential can be obtained exactly by solving the Schrödinger equation $H\psi(x) = E\psi(x)$ with
\begin{equation}\label{Morse-eq}
     H = -\frac{\hbar^2}{2m} \frac{d^2}{dx^2} + V(x)\,.
\end{equation}
Changing variables to $y = e^{-a(x - x_0)}$ and defining the dimensionless energy $\epsilon_n \equiv -\left(\lambda - n - \frac{1}{2} \right)^2$ puts \eqref{Morse-eq} into the form,
\begin{eqnarray}
    y^2 \frac{d^2 \psi}{dy^2} + y \frac{d \psi}{dy} + \left( -\frac{\epsilon_n}{(a^2)} + \frac{2\lambda^2}{y} - \frac{\lambda^2}{y^2} \right)\psi = 0\,,
\end{eqnarray}
where $\lambda = \frac{\sqrt{2m D_e}}{\hbar a}$ is a dimensionless parameter. A further definition of $\psi(y) = y^{s} e^{-y/2} \phi(y)$, with $s = \lambda - n - \frac{1}{2}$ reduces the Schr\"odinger equation to 
the associated Laguerre differential equation,
\begin{eqnarray}
    y \frac{d^{2}\phi(y)}{dy^{2}} + (\alpha + 1 - y)\frac{d\phi(y)}{dy} + n \phi(y) = 0\,,
\end{eqnarray}
with $\alpha = 2\lambda-2n-1$. The normalized wavefunctions are then given by
\begin{eqnarray}
    \psi_n(x) = \mathcal{N}_n \; y^{\lambda - n - \frac{1}{2}} e^{-y/2} L_n^{2\lambda - 2n - 1}(y)\,,
\end{eqnarray}
where 
\begin{eqnarray}
    L_{n}^{(\alpha )}(z)&=&{\frac {\ z^{-\alpha }\ e^{z}\ }{n!}}\ {\frac {\operatorname {d} ^{n}}{{\operatorname {d} z}^{n}}}\left(z^{n+\alpha }e^{-z}\right)\nonumber\\
    &=&{\frac {\ \Gamma (\alpha +n+1)\ \Gamma (\alpha +1)\ }{n!}}\;{}_{1}F_{1}(-n,\alpha +1,z)~.
\end{eqnarray}
is a generalized Laguerre polynomial and the normalization constant
\begin{eqnarray}
    \mathcal{N}_n = \left[ \frac{a (2\lambda - 2n - 1) \, \Gamma(n+1)}{\Gamma(2\lambda - n)} \right]^{1/2}\,.
\end{eqnarray}
The resulting quantized energy levels are given by
\begin{equation}
    E_n = \hbar\omega \left(n + \frac{1}{2}\right) - \frac{\hbar^2 a^2}{2m} \left(n + \frac{1}{2}\right)^2, \qquad n = 0, 1, \ldots, n_{\max},
    \label{eq:morsespectrum1}
\end{equation}
where
\begin{eqnarray}
    \omega = a \sqrt{\frac{2 D_e}{m}}, \qquad n_{\max} = \left\lfloor \frac{\sqrt{2m D_e}}{\hbar a} - \frac{1}{2} \right\rfloor\,,
\end{eqnarray}
and $\lfloor x \rfloor$ denotes the largest integer smaller than $x$. A few points are in order. First, the spectrum is discrete and bounded from below. This leads to a natural cutoff to the Hilbert space and makes the spectral sum in the microcanonical and thermal OTOC expressions convergent and computationally tractable. Unlike the harmonic oscillator, the Morse potential has decreasing level spacing as the energy increases. This leads to non-uniform phase accumulation in the spectral representation of \cite{Hashimoto:2017oit}, which clearly demonstrates how the time structure of OTOCs encodes spectral information. Second, the number of bound states is finite. This matches the intuition that the Morse potential has a continuum of scattering states for $E > 0$, and a finite number of bound states for $E < 0$. Third, none of this is new but it does lend itself to several interesting nuances. For example, the position operator connects multiple energy levels with amplitudes that depend nonlinearly on the quantum number. This makes the spectral representation of the OTOC sensitive to the full band structure, enabling tests of how scrambling builds up over finite-dimensional Hilbert spaces.\\

\begin{figure}[!htbp]

\centering
\subfloat[$D_e=2\Rightarrow N_{\text{max}}=3$\label{fig:de1}]{\includegraphics[width=7cm]{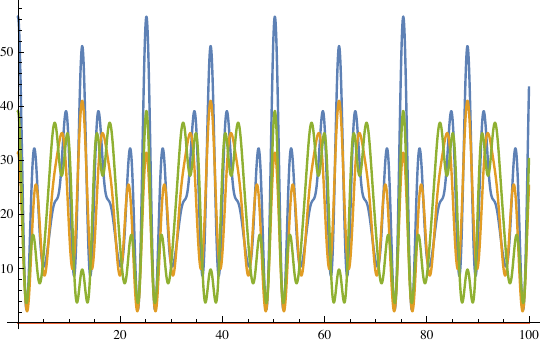}}\hfil
\subfloat[$D_e=2\Rightarrow N_{\text{max}}=3$\label{fig:de2}]{\includegraphics[width=7cm]{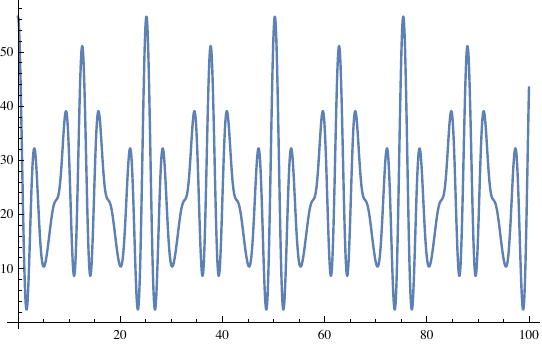}}\hfil 

\subfloat[$D_e=6\Rightarrow N_{\text{max}}=6$\label{fig:de3}]{\includegraphics[width=7cm]{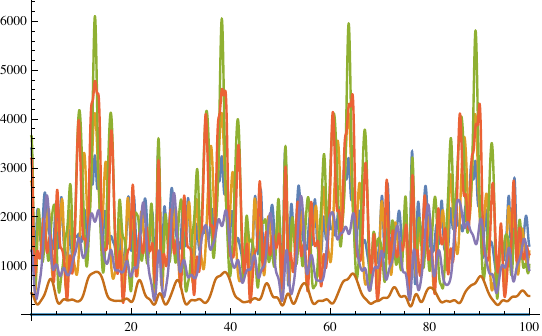}}\hfil
\subfloat[$D_e=6\Rightarrow N_{\text{max}}=6$\label{fig:de4}]{\includegraphics[width=7cm]{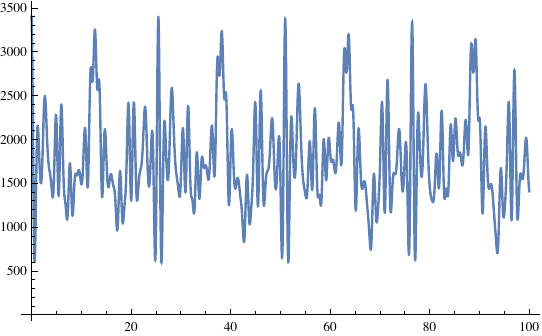}}\hfil 

\subfloat[$D_e=10\Rightarrow N_{\text{max}}=8$\label{fig:de5}]{\includegraphics[width=7cm]{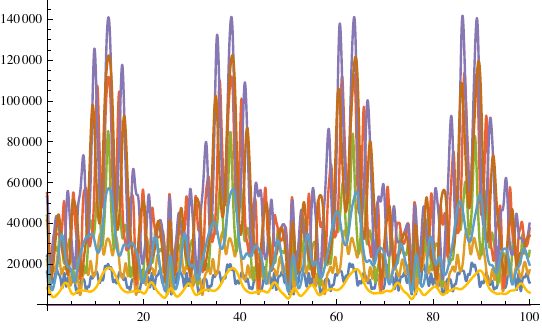}}\hfil
\subfloat[$D_e=10\Rightarrow N_{\text{max}}=8$\label{fig:de6}]{\includegraphics[width=7cm]{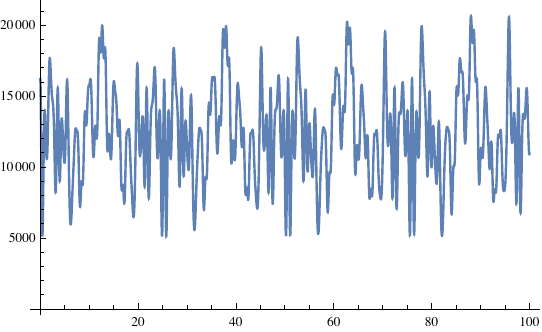}}\hfil 

\subfloat[$D_e=16\Rightarrow N_{\text{max}}=10$\label{fig:de7}]{\includegraphics[width=7cm]{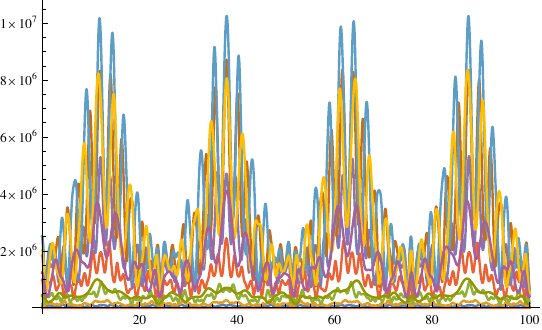}}\hfil
\subfloat[$D_e=16\Rightarrow N_{\text{max}}=10$\label{fig:de8}]{\includegraphics[width=7cm]{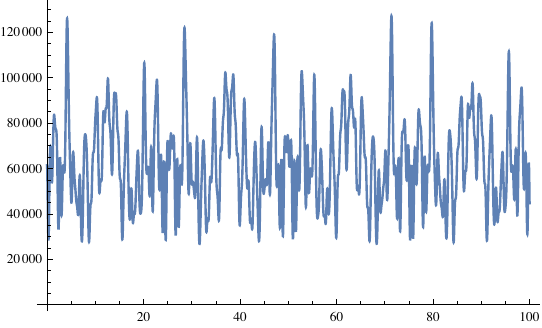}}\hfil 

\caption{Microcanonical (left) and thermal (right) OTOC's for $T=0.01$, $a=0.5$, $m=1$, and varying $D_e$ values.}
\label{figure:morseDepth}
\end{figure}

\begin{figure}[!htbp]

\centering
\subfloat[$a=0.3\Rightarrow N_{\text{max}}=18$\label{fig:a1}]{\includegraphics[width=7cm]{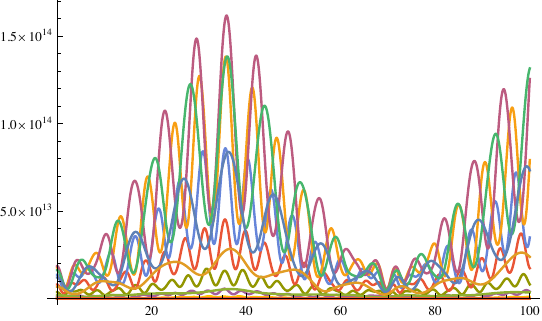}}\hfil
\subfloat[$a=0.3\Rightarrow N_{\text{max}}=18$\label{fig:a2}]{\includegraphics[width=7cm]{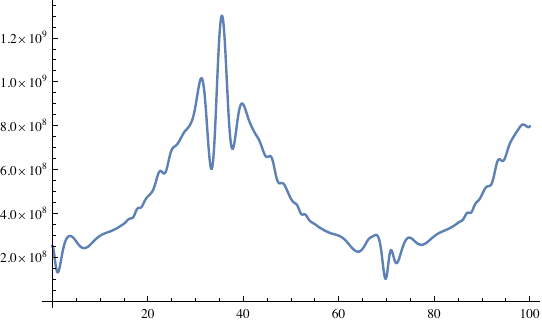}}\hfil 

\subfloat[$a=0.7\Rightarrow N_{\text{max}}=7$\label{fig:a3}]{\includegraphics[width=7cm]{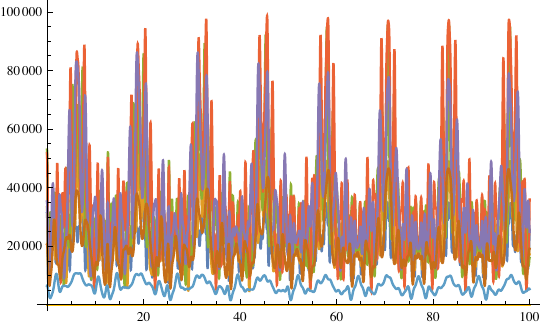}}\hfil
\subfloat[$a=0.7\Rightarrow N_{\text{max}}=7$\label{fig:a4}]{\includegraphics[width=7cm]{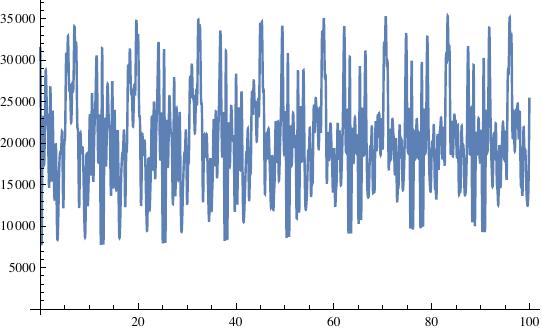}}\hfil 

\subfloat[$a=1.1\Rightarrow N_{\text{max}}=4$\label{fig:a5}]{\includegraphics[width=7cm]{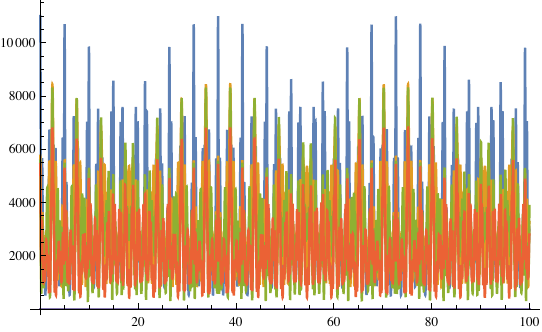}}\hfil
\subfloat[$a=1.1\Rightarrow N_{\text{max}}=4$\label{fig:a6}]{\includegraphics[width=7cm]{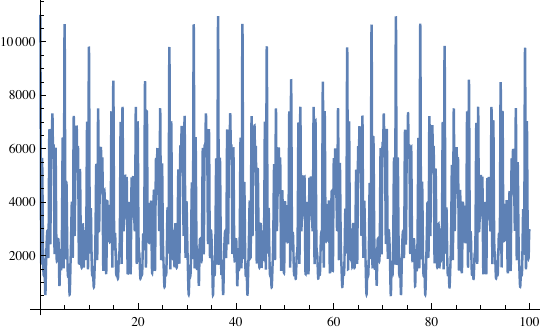}}\hfil 

\subfloat[$a=1.5\Rightarrow N_{\text{max}}=3$\label{fig:a7}]{\includegraphics[width=7cm]{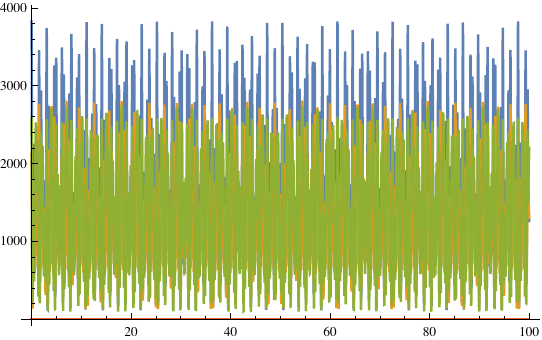}}\hfil
\subfloat[$a=1.5\Rightarrow N_{\text{max}}=3$\label{fig:a8}]{\includegraphics[width=7cm]{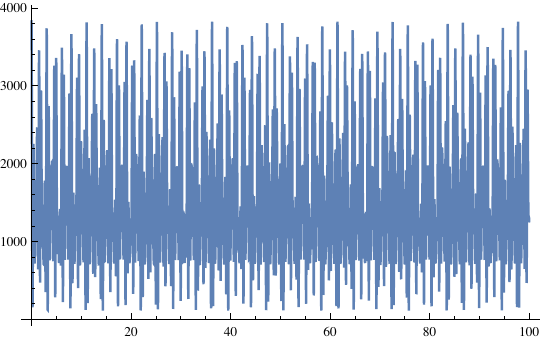}}\hfil 

\caption{Microcanonical (left) and thermal (right) OTOC's for $T=1.0$, $D_e=16$, $m=1$, and varying $a$ values.}
\label{figure:morseWidth}
\end{figure}

\begin{figure}[!htbp]

\centering
\subfloat[$T=0.01\Rightarrow N_{\text{max}}=8$]{\includegraphics[width=7cm]{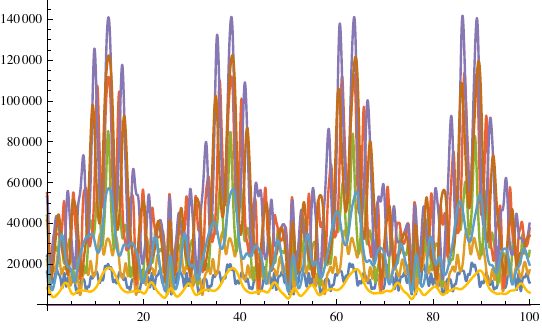}}\hfil
\subfloat[$T=0.01,$ $ N_{\text{max}}=8$]{\includegraphics[width=7cm]{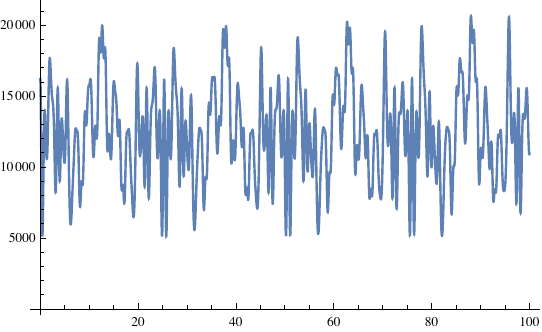}}\hfil 

\subfloat[$T=1,$ $ N_{\text{max}}=8$]{\includegraphics[width=7cm]{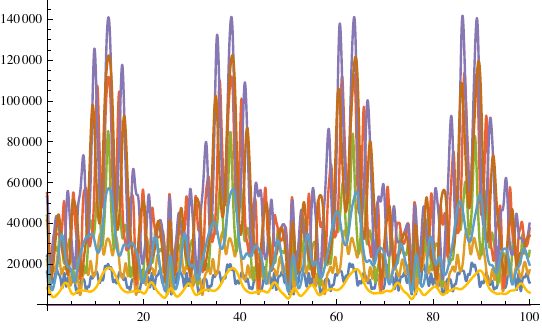}}\hfil
\subfloat[$T=1$, $N_{\text{max}}=8$]{\includegraphics[width=7cm]{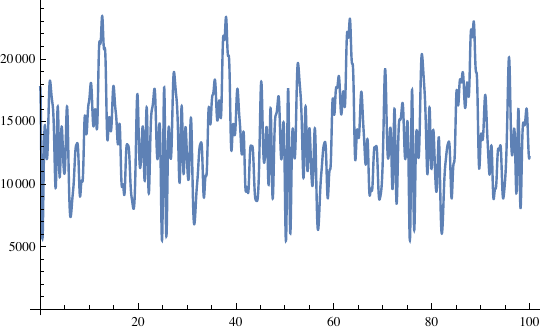}}\hfil 

\subfloat[$T=10,$ $N_{\text{max}}=8$]{\includegraphics[width=7cm]{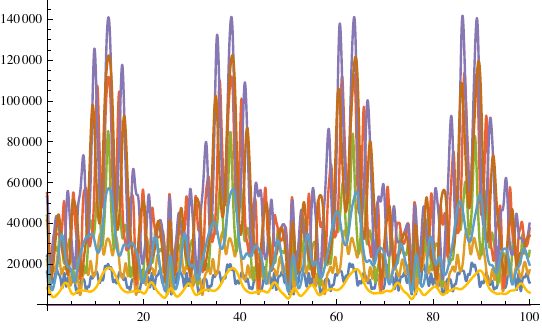}}\hfil
\subfloat[$T=10,$ $ N_{\text{max}}=8$]{\includegraphics[width=7cm]{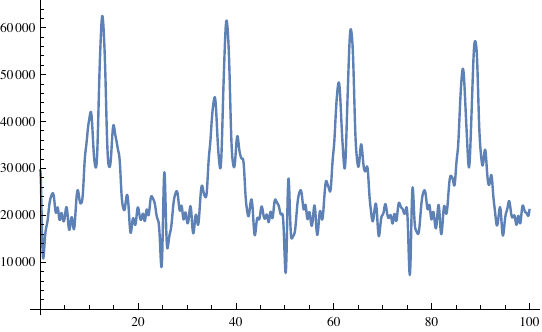}}\hfil 

\caption{Microcanonical (left) and thermal (right) OTOC's for $a=0.5$, $D_e=10$,  $m=1$, and varying $T$ values.}
\label{figure:morseTemp}
\end{figure}

\noindent
The matrix elements of the position operator in the Morse potential is quite a bit more complicated than the harmonic oscillator, but still known in closed-form. Specifically, the off diagonal elements are given by \cite{Lima_2005} as
\begin{eqnarray}
 x_{mn} \equiv \langle m|x|n\rangle =  \frac{2(-1)^{m - n + 1}}{(m - n)(2N - n - m)}
\sqrt{\frac{(N - n)(N - m)\Gamma(2N - m + 1)m!}{\Gamma(2N - n + 1)n!}}\,,\,  
\label{eq:morseposition}
\end{eqnarray}
while the diagonal elements can be expressed in terms of the digamma function $\psi(z = \Gamma'(z)/\Gamma(z))$ as
\begin{eqnarray}
    x_{nn}\equiv\langle n | x | n \rangle = \ln(2N + 1) + \psi(2N - n + 1) - \psi(2N - 2n + 1) - \psi(2N - 2n)\,.\,
\end{eqnarray}
These matrix elements, together with the known energies in \eqref{eq:morsespectrum1} are then substituted into \eqref{Bmn} which can then be used to determine the microcanonical, and subsequently, thermal OTOCs. In figures \ref{figure:morseDepth}-\ref{figure:morseTemp} we numerically evaluate $c_{n}(t)$ and $C_{T}(t)$ for various values of the potential depth, $D_{e}$, the potential width, $a$, and temperature, $T$. Overall, we observe that as $D_e$ increases, the number of bound states increases roughly as $N_{\text{max}} \sim \lambda = \sqrt{2m D_e}/(\hbar a)$. This leads to a much richer spectral interference and more pronounced temporal structure in $c_n(t)$. Some specific observations that deserve mention here are:
\begin{itemize}
    \item At low $D_e$ (Figs. \ref{fig:de1}-\ref{fig:de2}) only few states contribute, resulting in simple, low-amplitude, quasi-periodic oscillations.
	\item At intermediate $D_e$ (Figs. \ref{fig:de3}-\ref{fig:de6}), beat-like structures emerge, reflecting interference between multiple closely spaced levels. The OTOC captures resonances among these levels.
	\item At high $D_e$ ({\it i.e.} for deep wells, Figs. \ref{fig:de7}-\ref{fig:de8}) the dynamics resemble that of the harmonic oscillator at low energies, but the OTOCs exhibit more intricate modulation due to the anharmonicity.
    \item At small $a$: The well is broad and shallow, leading to fewer bound states and slow dynamics. The OTOC shows long timescales and low complexity in Fig. \ref{fig:a2}.
	\item At intermediate $a$: More levels and increased anharmonicity introduce richer oscillatory structure (see Figs. \ref{fig:a3}-\ref{fig:a6}). Peaks begin to show multi-periodic envelopes. 
	\item At large $a$: The well becomes steep and narrow—closer to a hard-wall limit. The level spacing increases and becomes more nonlinear. $c_n(t)$ exhibits rapid oscillations, quasi-periodicity, and a form of dephasing due to enhanced level dispersion, shown in Figs. \ref{fig:a7}-\ref{fig:a8}.
\end{itemize}
All in all, these results affirm that OTOCs are quite sensitive to the spectral density and level spacing of the system. Unlike in the harmonic oscillator however, Morse OTOCs show characteristic decay and revival patterns modulated by the potential’s anharmonicity. Finally, the contrast between low and high $T$ behavior (see Fig. \ref{figure:morseTemp}) highlights how thermal decoherence flattens out the time structure of the OTOC.

\section{OTOCs for a quantum particle in a magnetic field}
\label{sec:b-field_otoc}
The construction of microcanonical and thermal OTOCs proposed in Ref.~\cite{Hashimoto:2017oit} relies fundamentally on the spectral decomposition of the system: given a Hamiltonian $H$, one computes the eigenvalues $E_n$ and eigenfunctions $|n\rangle$, together with the matrix elements of the position operator $\hat{x}$. The OTOC can then be assembled from these ingredients via the quantities
\begin{equation}
b_{nm}(t) = \frac{1}{2}\sum_{k} x_{nk} x_{km}
\left[(E_k - E_m)e^{i(E_n - E_k)t} - (E_n - E_k)e^{i(E_k - E_m)t}\right],
\end{equation}
with the microcanonical OTOC given by
\begin{equation}
c_n(t) = \sum_m |b_{nm}(t)|^2, \qquad
C_T(t) = \frac{1}{Z}\sum_n e^{-\beta E_n} c_n(t).
\end{equation}
In the presence of a constant, transverse magnetic field $B$, this spectral construction remains valid, but its implementation requires modification at two key steps: (i) the determination of the eigenbasis, and (ii) the computation of position matrix elements.\\

\noindent
The Hamiltonian for a charged particle in two dimensions subject to a perpendicular magnetic field is
\begin{equation}
H = \frac{1}{2m} \left( \bm{p} - q \bm{A} \right)^2 + V(\bm{r}),
\label{mag-ham}
\end{equation}
with $V(\bm{r})$ enforcing the boundary conditions of the billiard. In the symmetric gauge, $\bm{A} = \tfrac{B}{2}(-y,x)$, this operator takes the explicit form
\begin{equation}
H = -\frac{\hbar^2}{2m}\left(\partial_x^2 + \partial_y^2\right)
	+	\frac{i\hbar q B}{2m} \left( y \partial_x - x \partial_y \right)
	+	\frac{q^2 B^2}{8m}(x^2+y^2) + V(x,y).
\end{equation}
The eigenstates of this operator may be computed by finite-element or finite-difference methods, or by expansion in Landau-level wavefunctions with boundary matching on the billiard. The resulting set of eigenpairs $\{E_n, \psi_n(x,y)\}$ provide the required input for the OTOC construction. A crucial difference from the zero-field case is that in the presence of a magnetic field the canonical momentum $\bm{p}$ does not commute with $\bm{A}$, and position operators become intertwined with the gauge choice. However, as we have seen above, the OTOC is expressible purely in terms of the matrix elements of the position operator,
\begin{equation}
   x_{nm} = \int_{\Omega} d^2r\, \psi_n^*(x,y)\,x\,\psi_m(x,y)\,.
\end{equation}
This integral is well defined independently of gauge, so long as one consistently uses the eigenfunctions $\psi_n(x,y)$ obtained in a chosen gauge. This means that the algorithmic structure of the OTOC remains intact, with the only new challenge being the accurate numerical evaluation of these integrals in the presence of magnetic confinement and edge-localized states.\\

\noindent
From this perspective, the extension of the construction in \cite{Hashimoto:2017oit} to magnetic billiards is natural. Nevertheless, it has important physical consequences. At large B, the spectrum organizes into Landau levels modified by the billiard boundary, producing edge states with reduced spatial extent. The position operator couples Landau-like states differently from plane-wave states, leading to suppressed long-range contributions in $x_{nm}$. The early-time quadratic growth of OTOCs remains universal, but the coefficient decreases as $B$ grows, reflecting the reduced spatial extent of wavefunctions. At late times, the saturation value of the thermal OTOC is also suppressed relative to the zero-field case, consistent with the localization of wavefunctions into cyclotron orbits and edge states. In chaotic billiards, the addition of a magnetic field does not generate a clear exponential regime in the OTOC; rather, it modifies the amplitude, frequency content, and saturation value of the oscillatory signal. This framework therefore provides a controlled way to quantify the impact of transverse magnetic fields on quantum information scrambling.

\subsection{Worked example: circular magnetic billiard}
\begin{figure}[!htbp]

\centering
\subfloat[Disk; $c_n(t)$; $B=0.0$\label{fig:disk1}]{\includegraphics[width=7cm]{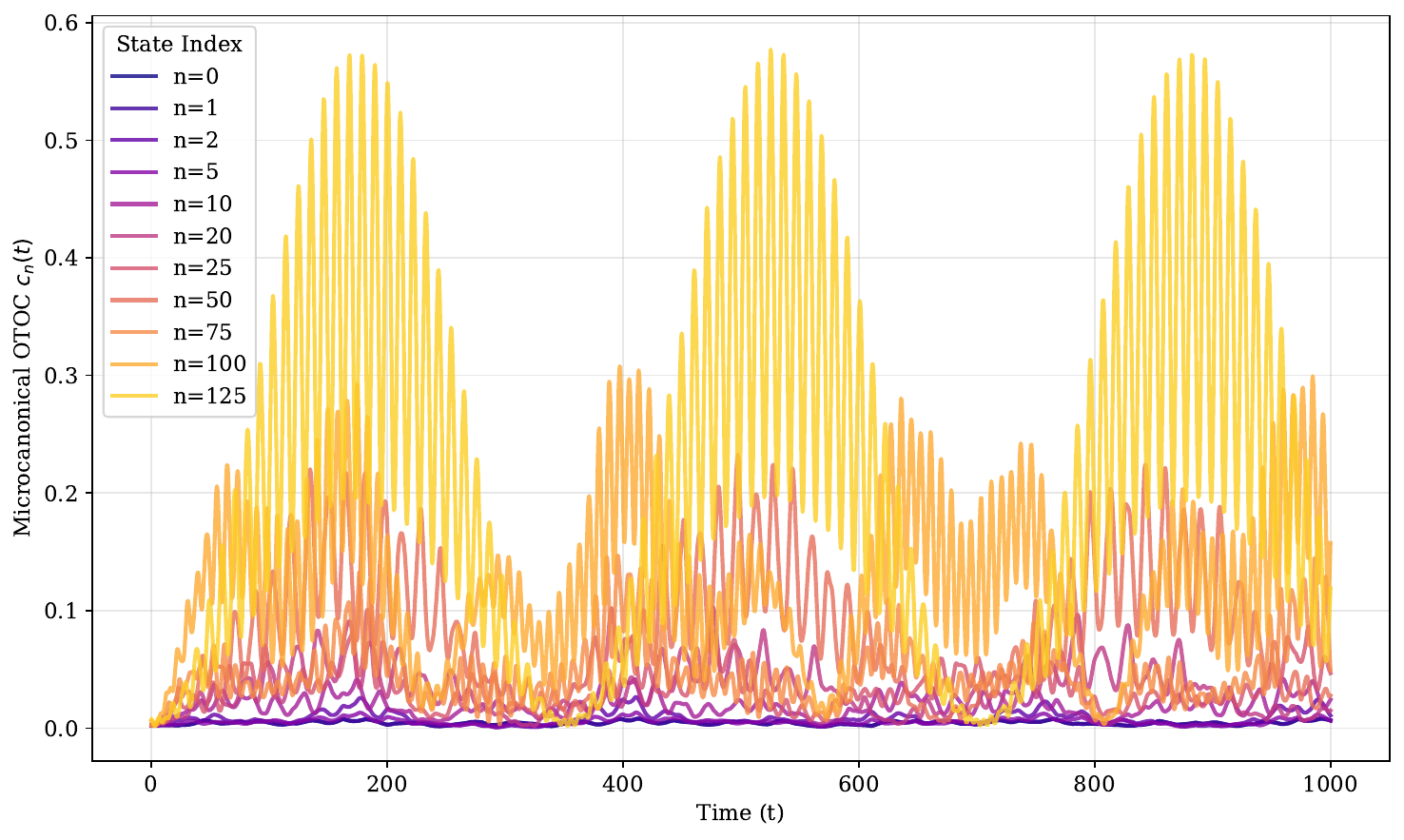}}\hfil
\subfloat[Disk; $C_T(t)$; $B=0.0$\label{fig:disk2}]{\includegraphics[width=7cm]{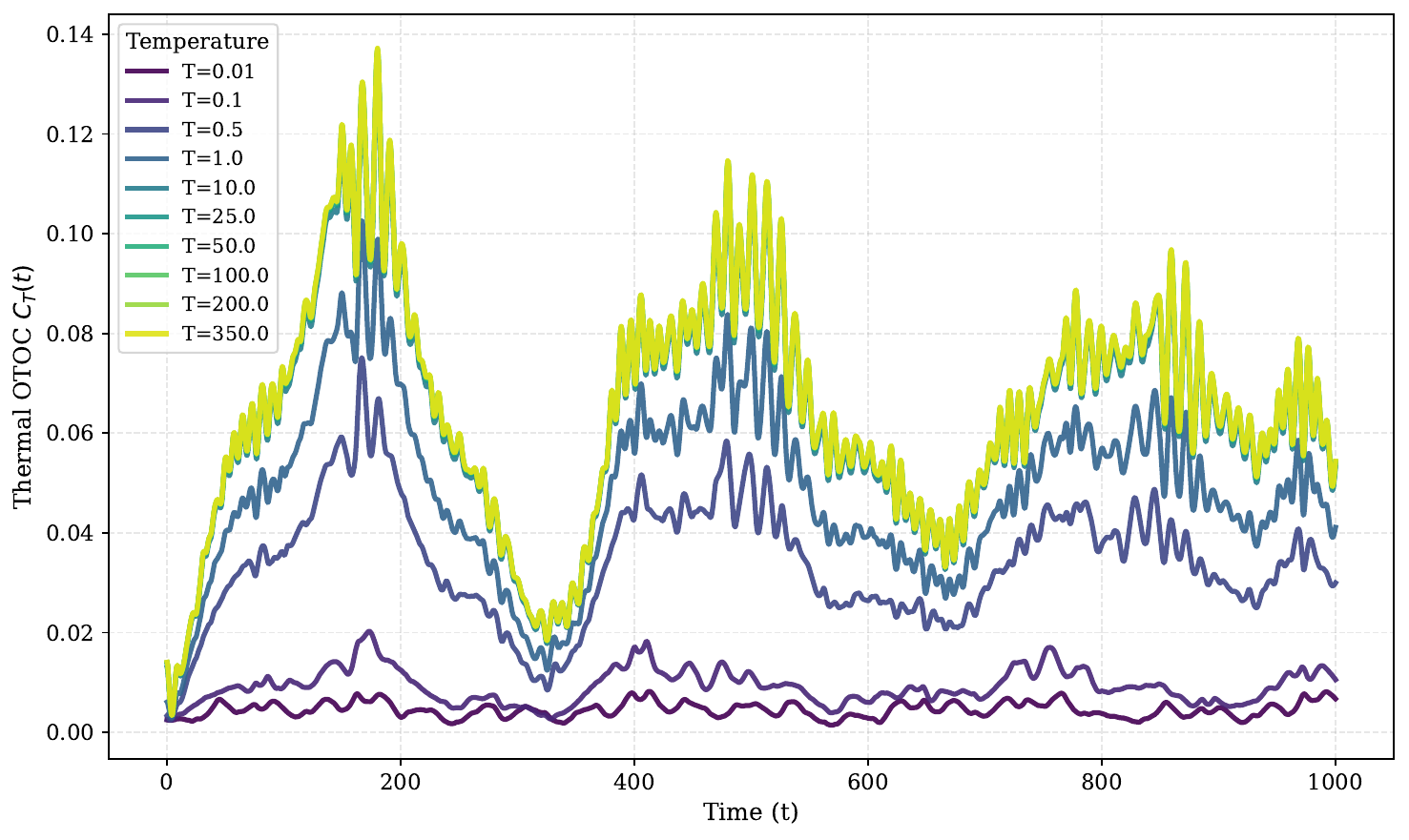}}\hfil 

\subfloat[Disk; $c_n(t)$; $B=1.0$\label{fig:disk3}]{\includegraphics[width=7cm]{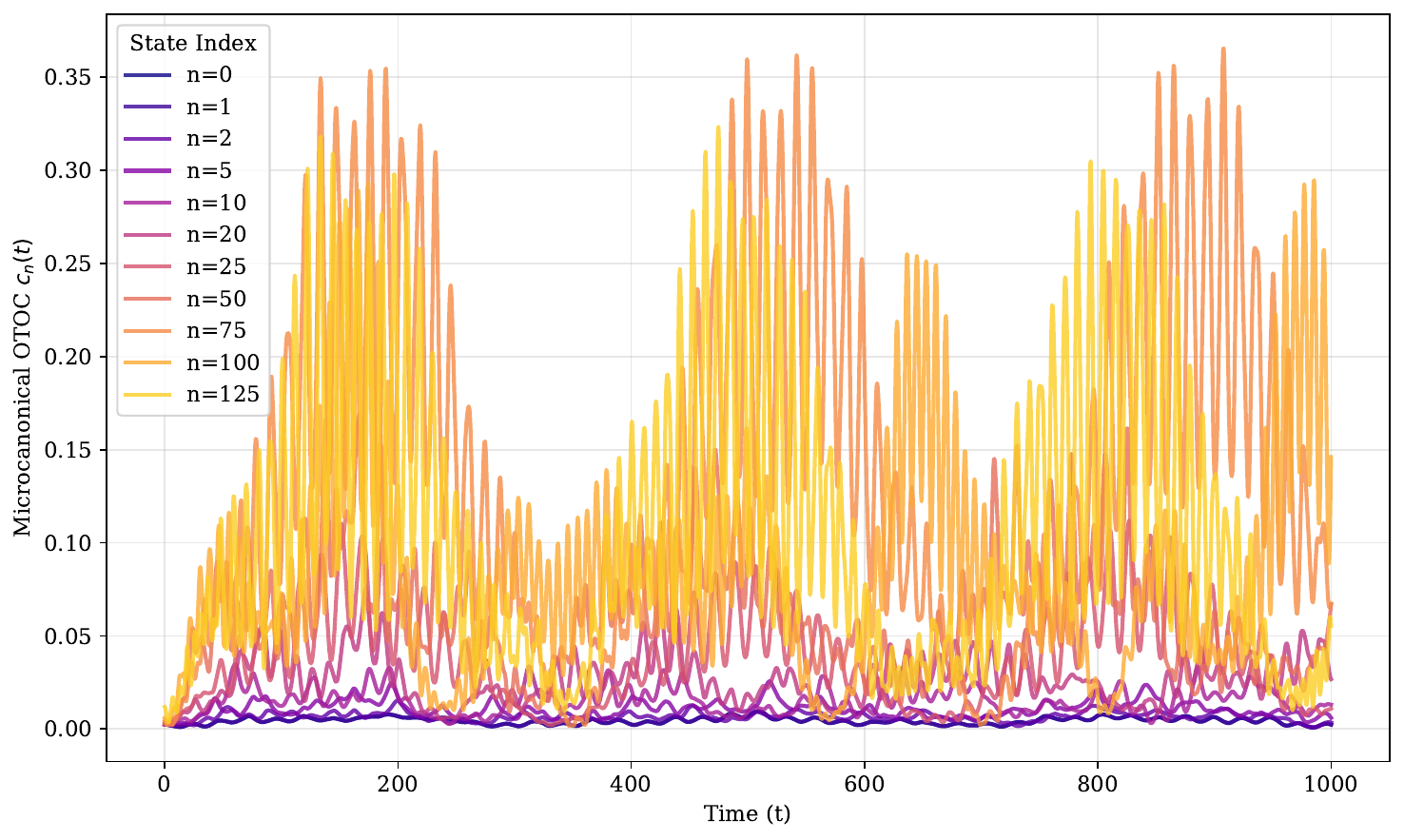}}\hfil
\subfloat[Disk; $C_T(t)$; $B=1.0$\label{fig:disk4}]{\includegraphics[width=7cm]{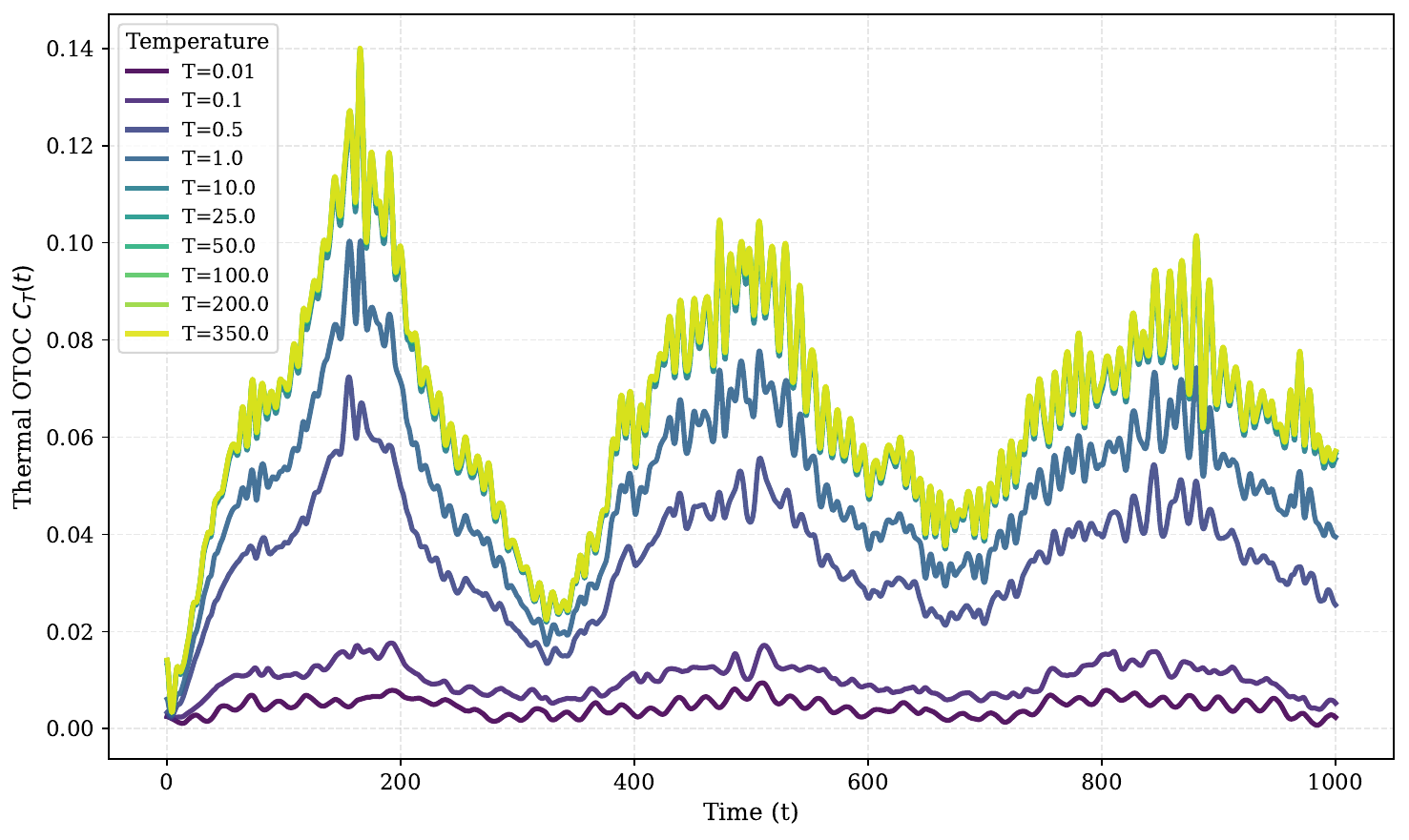}}\hfil 

\subfloat[Disk; $c_n(t)$; $B=2.0$\label{fig:disk5}]{\includegraphics[width=7cm]{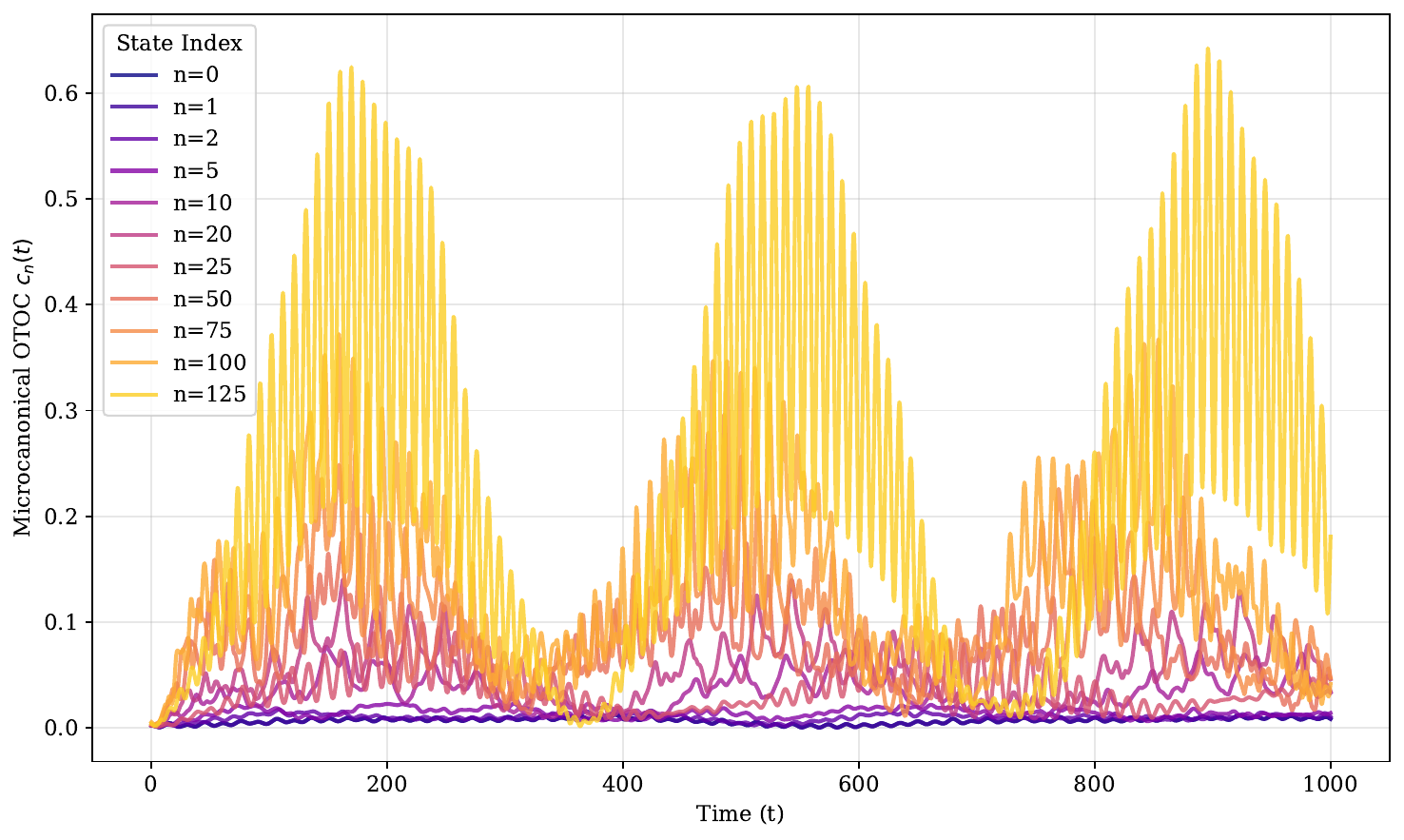}}\hfil
\subfloat[Disk; $C_T(t)$; $B=2.0$\label{fig:disk6}]{\includegraphics[width=7cm]{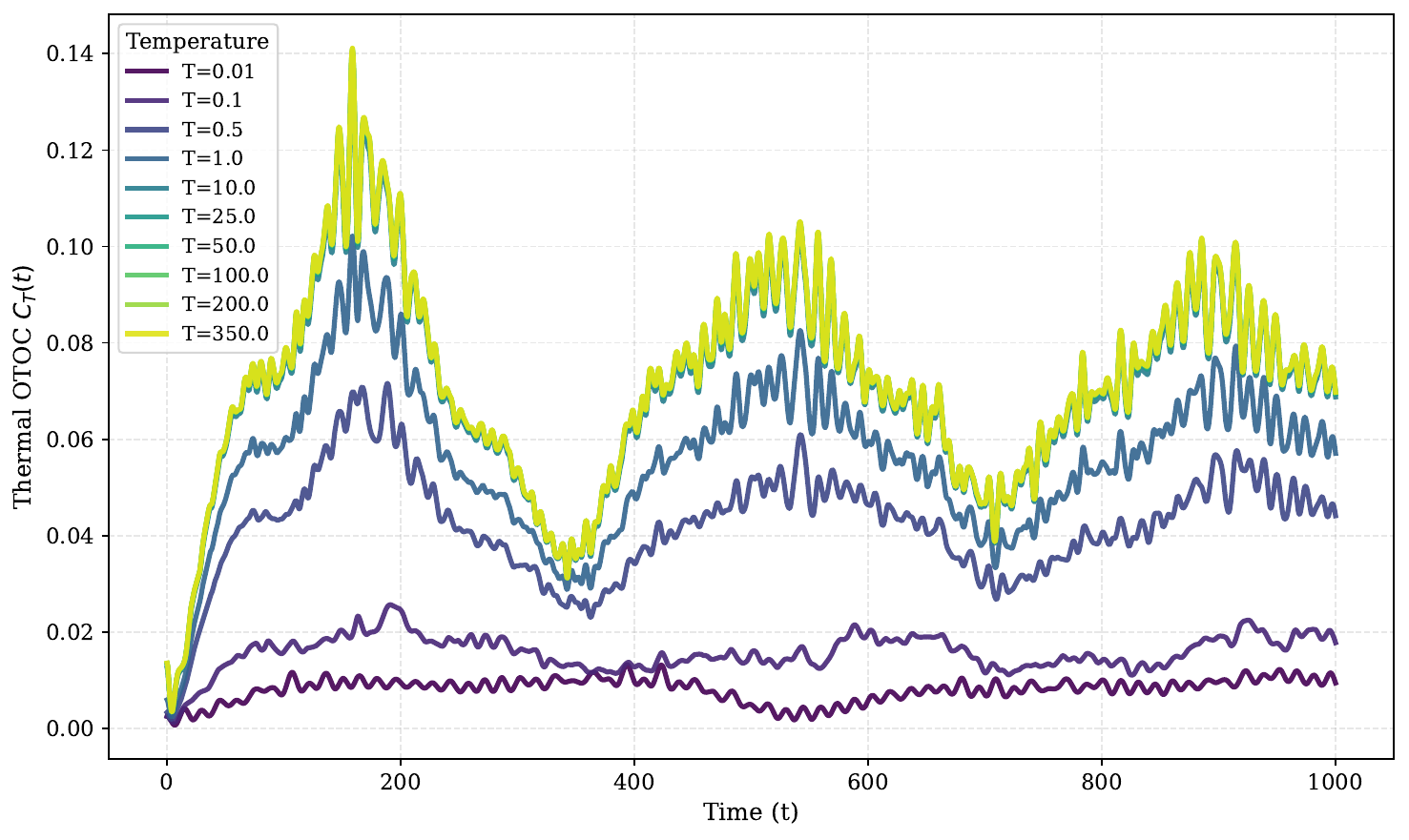}}\hfil 

\subfloat[Disk; $c_n(t)$; $B=4.0$\label{fig:disk7}]{\includegraphics[width=7cm]{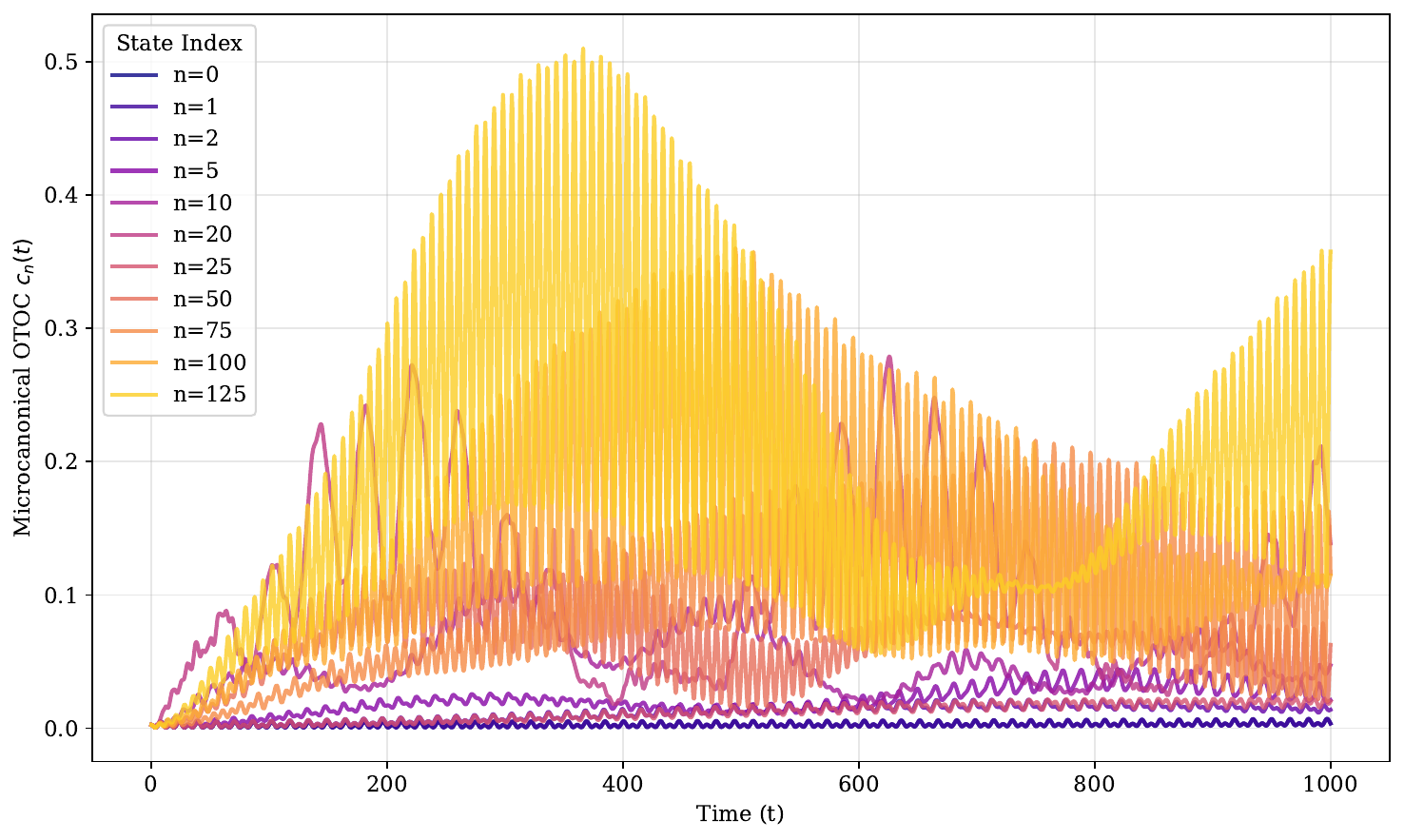}}\hfil
\subfloat[Disk; $C_T(t)$; $B=4.0$\label{fig:disk8}]{\includegraphics[width=7cm]{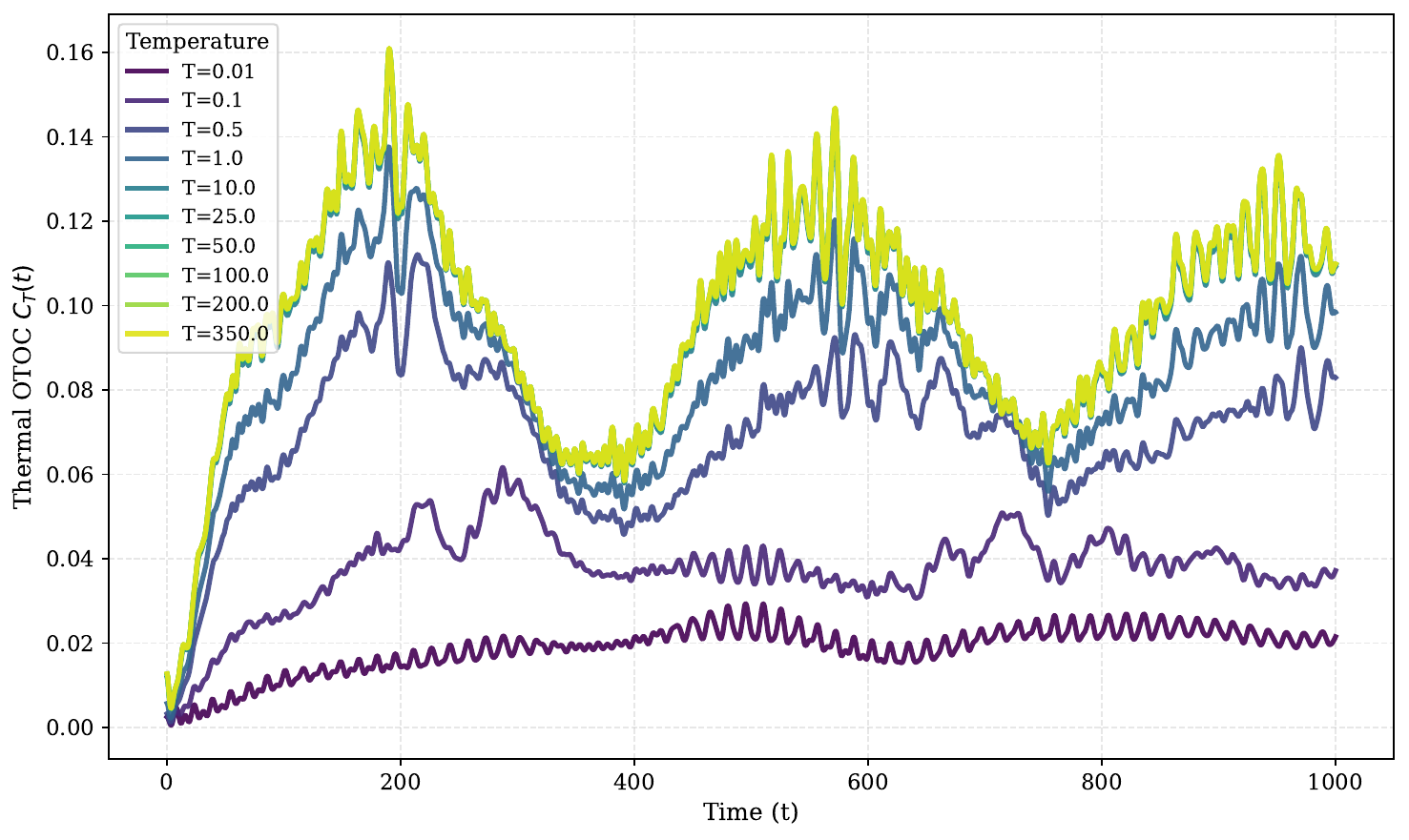}}\hfil

\caption{Microcanonical (left) and thermal (right) OTOC's for a charged particle in a circular magnetic billiard $B=0, 1, 2$ and $4$.}
\label{figure:magnetic-circle}
\end{figure}

As a concrete illustration of these ideas, let's now consider a charged particle of mass $m$ and charge $q$ confined to a disk of radius $R$ in a uniform magnetic field $B\hat{\bm{z}}$. In the symmetric gauge,
\begin{eqnarray}
 \bm{A}=\frac{B}{2}(-y,x),\qquad
 H=\frac{1}{2m}\bigl(-i\hbar\bm{\nabla}-q\bm{A}\bigr)^2,
 \quad \psi|_{r=R}=0.
 \label{symmetric-gauge-ham}
\end{eqnarray}
Define the magnetic length $b\equiv\sqrt{\frac{2\hbar}{qB}}$ (and take $qB>0$ for definiteness). In polar coordinates $(r,\theta)$, the bulk solutions of $H$ (before imposing the boundary) can be written as
\begin{align}
\psi_{a,m}(r,\theta)
&= \mathcal N_{a,m}\;e^{im\theta}\left(\frac{r}{b}\right)^{|m|}
\exp\left(-\frac{r^2}{2b^2}\right)\,
{}_1F_1\!\left(-a;|m|+1;\frac{r^2}{b^2}\right),
\label{eq:bulkpsi}
\end{align}
with integer angular momentum $m\in\mathbb{Z}$ and a non-negative real parameter $a$ that plays the role of a radial quantum number. Dirichlet boundary conditions on the circle fix $a$ by the quantization condition
\begin{equation}
   {}_1F_1\!\left(-a;|m|+1;\frac{R^2}{b^2}\right)=0,
   \qquad a\equiv a_{n_r,m},\quad n_r=0,1,2,\dots,
   \label{eq:rootcond}
\end{equation}
which produces a discrete set of allowed values 
$\{a_{n_r,m}\}$ for each $m$. For each admissible pair $(n_r,m)$ the energy is
\begin{equation}
   E_{n_r,m}
   = \hbar\omega_c\left(a_{n_r,m}+\frac{m+|m|+1}{2}\right),
   \qquad \omega_c\equiv\frac{qB}{m}\,.
   \label{eq:Em}
\end{equation}
The normalization $\mathcal N_{a,m}$ is fixed by
\begin{equation}
   1=\int_{r<R}\!d^2r\,|\psi_{a,m}|^2
   =2\pi\,|\mathcal N_{a,m}|^2\int_0^R\!dr\,r\left(\frac{r}{b}\right)^{2|m|}
   e^{-r^2/b^2}\,\big|{}_1F_1(-a;|m|+1;r^2/b^2)\big|^2\,.
\label{eq:norm}
\end{equation}
Now let $|\alpha\rangle\equiv|n_r,m\rangle$ denote the normalized eigenstates. We need
$x_{\alpha\beta}\equiv\langle \alpha|x|\beta\rangle$
to build the OTOC. With $x=r\cos\theta=\tfrac{r}{2}(e^{i\theta}+e^{-i\theta})$ and the angular dependence $e^{im\theta}$, the angular integral enforces the selection rule
\begin{equation}
   \int_0^{2\pi}\!d\theta\, e^{i(m’-m)\theta}\cos\theta
   =\pi\big(\delta_{m’,m+1}+\delta_{m’,m-1}\big)\,.
   \label{eq:angsel}
\end{equation}
Hence,
\begin{eqnarray}
   x_{(n_r,m),(n’_r,m’)}
   &=& \pi\,\delta_{m’,m\pm1}\;
   \mathscr{I}_{(n_r,m)\to(n’r,m’)}\;,
   \nonumber\\
   \mathscr{I}_{(n_r,m)\to(n’_r,m’)}
   &=&\int_0^R dr\, r^2\, R_{n_r,m}(r),R_{n’_r,m’}(r),
   \label{eq:xmatrix}
\end{eqnarray}
where $R_{n_r,m}(r)$ is the radial part in \eqref{eq:bulkpsi} with the proper normalization. Writing $u\equiv r^2/b^2$ gives a single-integral representation convenient for numerics,
\begin{eqnarray}
   \mathscr{I}_{(n_r,m)\to(n’_r,m\pm1)}
   &=&\frac{b^{|m|+|m\pm1|+4}}{2}\,\mathcal N_{a_{n_r,m},m}\,\mathcal N_{a_{n’_r,m\pm1},m\pm1}
   \int_0^{u_R}\!du\; u^{\frac{|m|+|m\pm1|+1}{2}} e^{-u}
   \nonumber\\
   &\times&
   {}_1F_1\!\left(-a_{n_r,m};|m|+1;u\right)\;
   {}_1F_1\!\left(-a_{n’_r,m\pm1};|m\pm1|+1;u\right)\,,
   \label{eq:radialI}
\end{eqnarray}
with $u_R=R^2/b^2$. Eqs.~\eqref{eq:norm} and \eqref{eq:radialI} determine all position matrix elements $x_{\alpha\beta}$ with a single quadrature, since the angular factor is in closed-form. (In practice, a Gauss–Laguerre or adaptive quadrature on $u\in[0,u_R]$ yields rapid convergence.) Labeling eigenpairs by $\alpha=(n_r,m)$ and defining the energy gaps $E_{\alpha\beta}\equiv E_\alpha-E_\beta$ gives
\begin{equation}
b_{\alpha\beta}(t)
=\frac{1}{2}\sum_{\gamma}
x_{\alpha\gamma}\,x_{\gamma\beta}\,
\Big[(E_\gamma\!-\!E_\beta)e^{iE_{\alpha\gamma}t}
-(E_\alpha\!-\!E_\gamma)e^{iE_{\gamma\beta}t}\Big],
\label{eq:bab}
\end{equation}
which feeds directly into the microcanonical and thermal OTOCs via
\begin{equation}
   c_\alpha(t)=\sum_{\beta}|b_{\alpha\beta}(t)|^2,
   \qquad
   C_T(t)=\frac{1}{Z}\sum_{\alpha}e^{-\beta E_\alpha}\,c_\alpha(t)\,,
   \qquad Z=\sum_\alpha e^{-\beta E_\alpha}\,.
   \label{eq:ctherm}
\end{equation}
Since $x$ only couples $m\to m\pm1$ (Eq.~\eqref{eq:angsel}), the sum over $\gamma$ in \eqref{eq:bab} is sparse, which substantially reduces the cost of assembling $b_{\alpha\beta}$ and $c_\alpha$.
At this stage, it is worth carrying out some sanity checks on these results:
\begin{itemize}
    \item Zero magnetic field (Figs. \ref{fig:disk1}-\ref{fig:disk2}): The roots $a_{n_r,m}$ reorganize so that $\psi$ reduces to Bessel modes with zeros at $r=R$; \eqref{eq:xmatrix} reproduces the usual selection rule and radial Bessel overlaps. The OTOC matches the field-free circular billiard of \cite{Hashimoto:2017oit}.
    \item Large magnetic field (Figs. \ref{fig:disk7}-\ref{fig:disk8}): States cluster into boundary-dressed Landau ladders; radial weight shifts toward the edge. The overlaps $\mathscr I$ shrink in effective range, yielding smaller early-time curvature and a reduced late-time saturation in the time-averaged $C_T(t)$.
    \item Early-time coefficient: From $H=\Pi^2/2m$ with kinetic momentum $\bm{\Pi}=\bm p-q\bm A$, one finds $[x,H]=i\hbar\,\Pi_x/m\, \mathrm{and}\, [\, [x,H],H\,]=-(\hbar^2 q B/m^2)\,\Pi_y$. Thus the universal $t^2$ growth of the thermal OTOC has a $B$-dependent prefactor controlled by $\langle\Pi_y\rangle_T$, which decreases as cyclotron localization increases.
\end{itemize}
This example shows that, aside from the replacement of the Bessel basis by confluent-hypergeometric functions and a boundary root search, the spectral construction of \cite{Hashimoto:2017oit} carries over verbatim to magnetic billiards. The magnetic field, of course alters the spectra and the sparsity pattern of $x_{\alpha\beta}$, which in turn modulates the OTOC’s early-time curvature and late-time saturation in physically transparent ways. The results of our numerical computations are given in Fig. \ref{figure:magnetic-circle}.\\

\subsection{From circular to general billiards}

For non-separable boundary geometries such as the stadium or Sinai billiards, the spectral ingredients for the construction of OTOCs can no longer be computed analytically, in closed form. Fortunately, they are naturally provided by so-called {\it boundary methods} which reduce the computation of eigenvalues and wave functions to a problem involving only the boundary $\partial\mathscr{B}$. For billiards in a constant (transverse) magnetic field, these were computed some time ago by de Aguiar in \cite{PhysRevE.53.4555}.\\

\noindent
In de~Aguiar’s construction one solves the magnetic Schr\"odinger problem
\begin{equation}
   H=\frac{1}{2m}\bigl(-i\hbar\bm{\nabla}-q\bm {A}\bigr)^2,\qquad 
   \bm{A}=\tfrac{B}{2}(-y,x),\qquad \psi|_{\partial\mathscr{B}}=0,
\end{equation}
by expanding the wavefunction in a basis adapted to Landau dynamics and enforcing boundary conditions through a boundary integral equation.  For the disk this reproduces exactly the confluent–hypergeometric basis we used above; for a generic billiard $\mathscr{B}$ it yields \emph{numerical} eigenpairs $\{E_\alpha,\psi_\alpha(\bm r)\}$ that converge rapidly with basis size or boundary discretization.  These are precisely the inputs required by the OTOC formulas.
Given a basis $\psi_\alpha(\bm r)$, the position matrix elements follow from the domain integral
\begin{equation}
   x_{\alpha\beta}=\int_{\mathscr{B}}\! d^2r\,\psi_\alpha^*(\bm r)\;x\;\psi_\beta(\bm r)\,,
   \label{eq:x-matrix-generic}
\end{equation}
in the same gauge used to compute the eigenstates.\\

\noindent
Gauge consistency is crucial, but the quantity in \eqref{eq:x-matrix-generic} itself is gauge invariant because the phases cancel between bra and ket.  
In the circular case, angular momentum yields the selection rule $m\to m\pm1$, which sparsifies the sums in the OTOC coefficients. For chaotic geometries there is no such simple rule; numerically one still assembles the finite matrix $x_{\alpha\beta}$ by quadrature on a bulk grid or by reusing the boundary operator to evaluate overlaps. Either route drops directly into the building blocks of \cite{Hashimoto:2017oit},
\begin{equation}
   b_{\alpha\beta}(t)=\frac12\sum_{\gamma}
   x_{\alpha\gamma}x_{\gamma\beta}\Big[(E_\gamma-E_\beta)e^{i(E_\alpha-E_\gamma)t}
   -(E_\alpha-E_\gamma)e^{i(E_\gamma-E_\beta)t}\Big],\qquad
   c_\alpha(t)=\sum_\beta|b_{\alpha\beta}(t)|^2,
\end{equation}
and the thermal average $C_T(t)=Z^{-1}\sum_\alpha e^{-\beta E_\alpha}c_\alpha(t)$ computed. To summarize the algorithm, we:

\begin{enumerate}
\item Compute $\{E_\alpha,\psi_\alpha\}$ for the chosen billiard $\mathscr{B}$ and field $B$ with the  boundary scheme in \cite{PhysRevE.53.4555} (or an equivalent FEM/FD discretization) up to an energy cutoff $E_{\rm cut}$.
\item Form $x_{\alpha\beta}$ from \eqref{eq:x-matrix-generic}; in practice, adopt an interior quadrature mesh and validate by symmetry checks (parities at $B{=}0$, edge localization at large $B$).
\item Assemble $b_{\alpha\beta}(t)$ and $c_\alpha(t)$; monitor \emph{truncation stability} by increasing $E_{\rm cut}$ until the curves on the time window of interest have converged.
\end{enumerate}

\noindent
Three limits tie the numerical implementation to analytic expectations:
(i) $B\!\to\!0$ recovers the field-free billiards of~\cite{Hashimoto:2017oit}; 
(ii) the circular geometry reproduces the Kummer–function solution that we found above, including the $m{\to}m\pm1$ sparsity; 
(iii) $B$ large yields Landau-like ladders and edge/skipping states.  In this regime we find (and numerics confirm) a reduced early-time curvature of $C_T(t)$ and a suppressed long-time saturation, reflecting cyclotron localization and the shorter effective excursion length along the boundary.

\subsection{Thermal OTOCs for the magnetic stadium billiard}
\begin{figure}[!htbp]

\centering
\subfloat[Stadium; $C_T(t)$; $B=0.0$\label{fig:stadium1}]{\includegraphics[width=7cm]{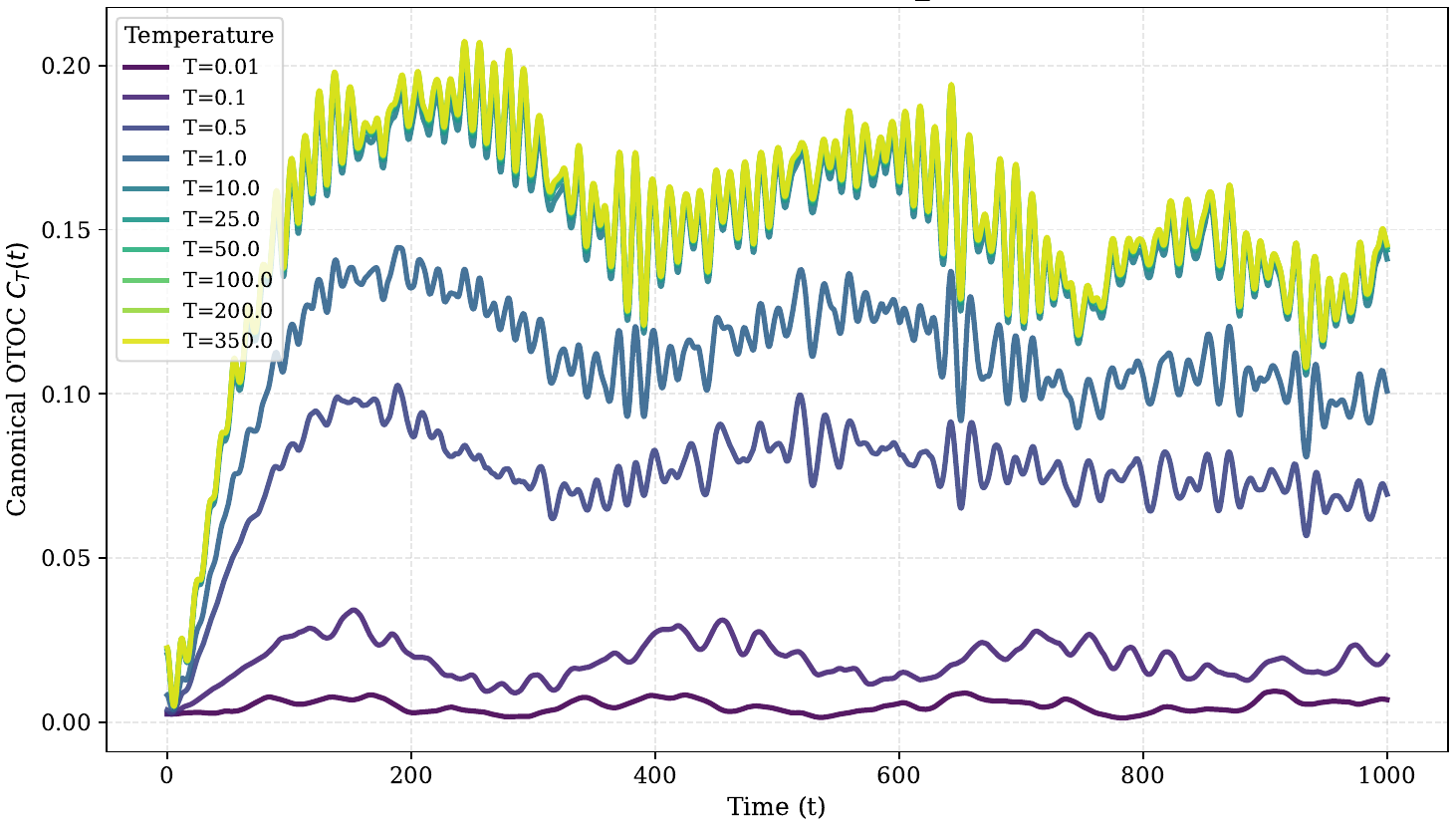}}\hfil
\subfloat[Stadium; $C_T(t)$; $B=1.0$\label{fig:stadium2}]{\includegraphics[width=7cm]{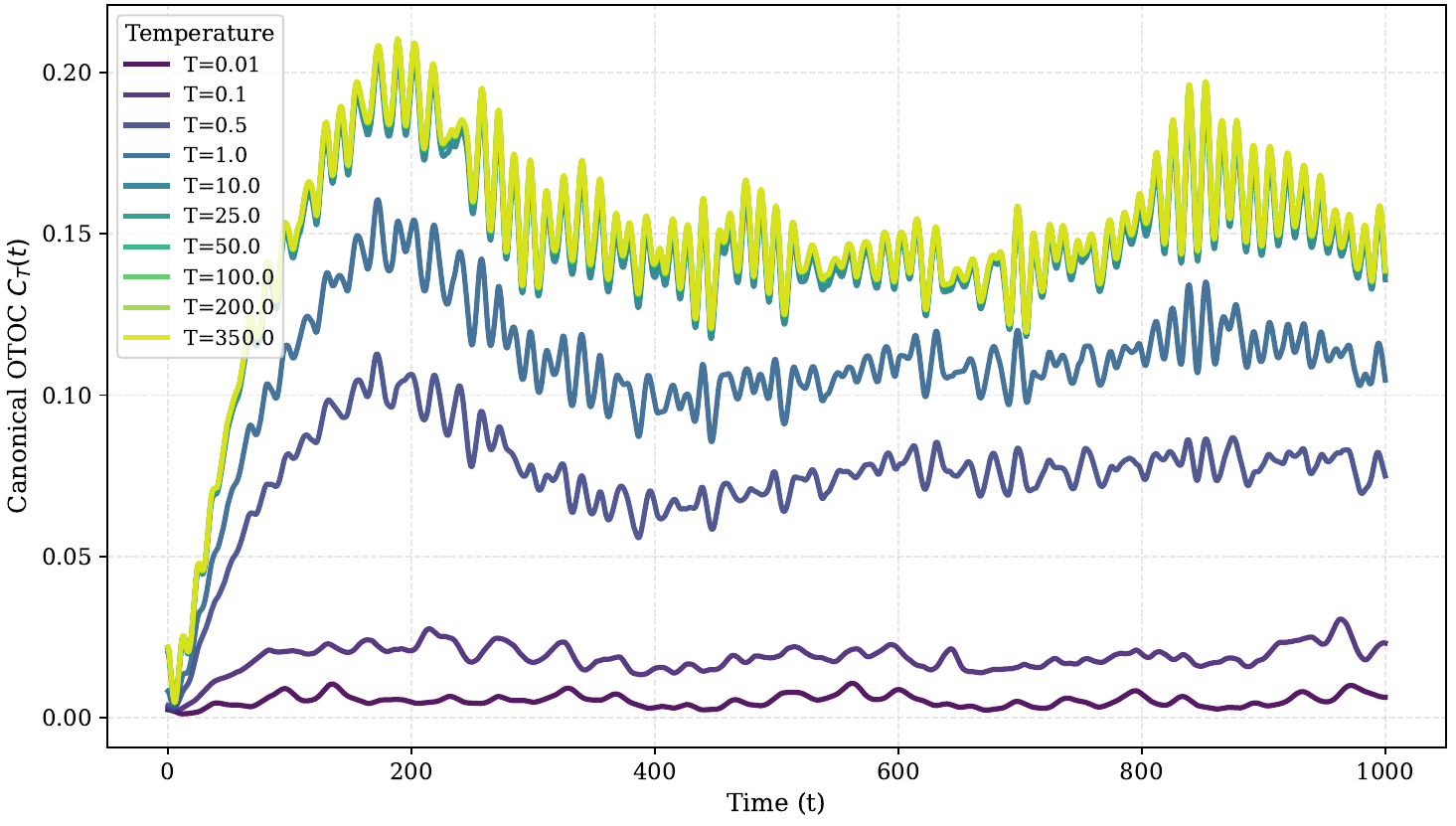}}\hfil 

\subfloat[Stadium; $C_T(t)$; $B=2.0$\label{fig:stadium3}]{\includegraphics[width=7cm]{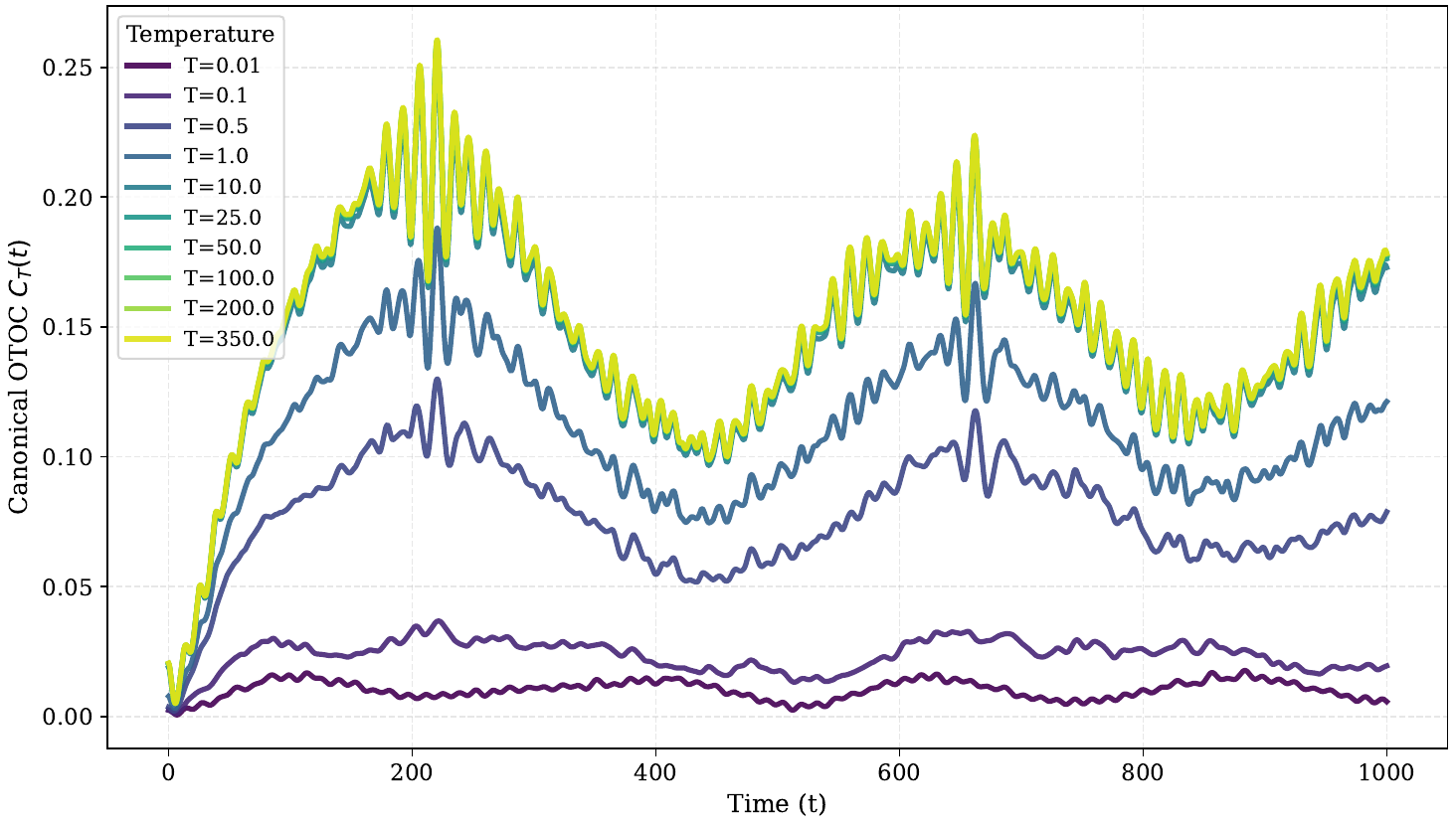}}\hfil
\subfloat[Stadium; $C_T(t)$; $B=4.0$\label{fig:stadium4}]{\includegraphics[width=7cm]{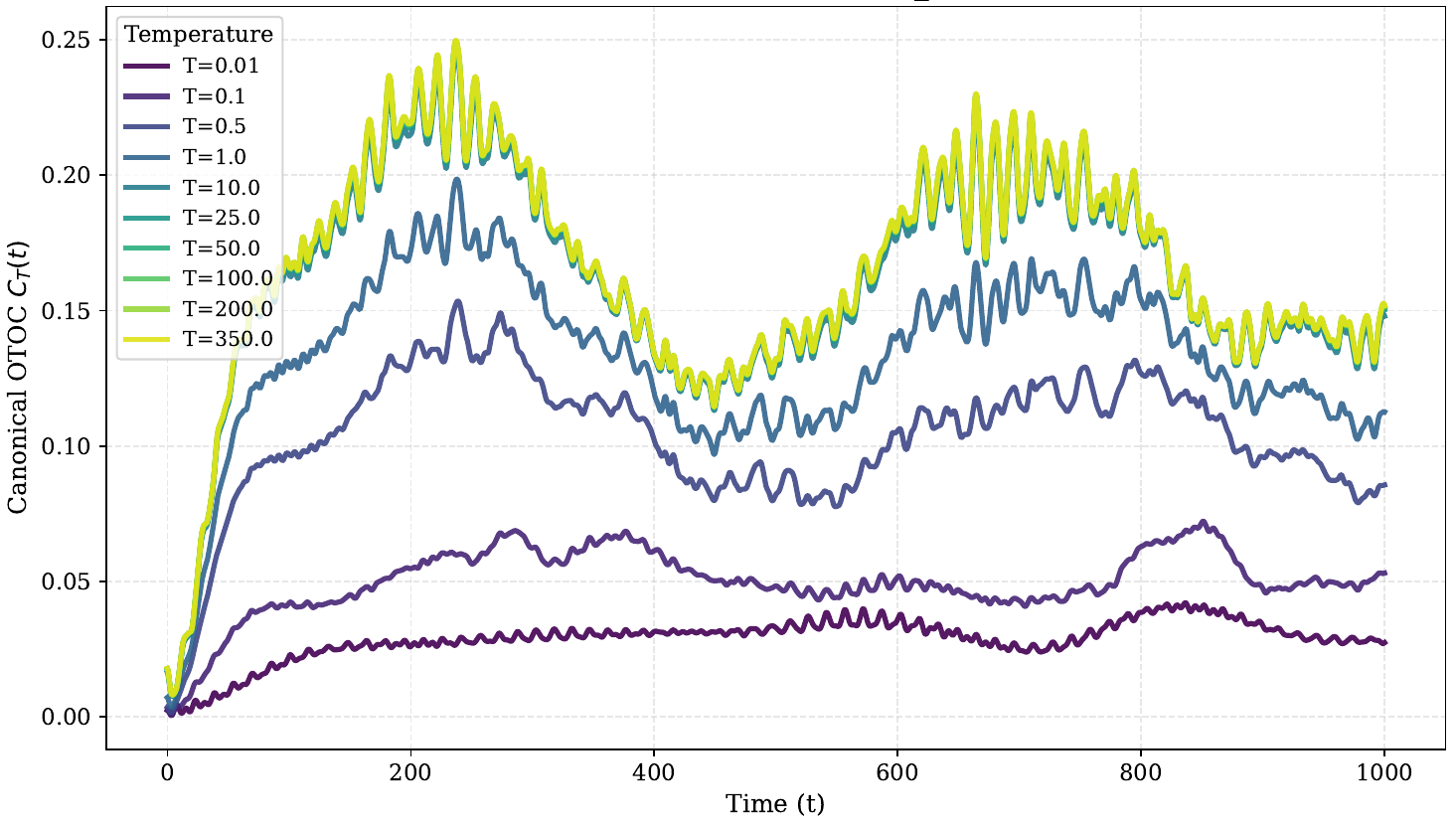}}\hfil 

\caption{Thermal OTOC's for a charged particle in a stadium magnetic billiard with magnetic field strengths $B=0, 1, 2$ and $4$.}
\label{figure:magnetic-stadium}
\end{figure}

To illustrate the OTOC construction above, we consider a charged particle in a two-dimensional stadium in which the particle is free within the stadium and reflects specularly off the boundary resulting in one of the most well-known classically chaotic systems, the stadium, or Bunimovich billiard. The confining potential for the billiard system
\begin{eqnarray*}
    V(x, y) = 
    \begin{cases}
        0 & \text{if } (x, y) \in \mathscr{D}_{\text{stadium}}, \\
        \infty & \text{otherwise},
    \end{cases}
    \quad \text{with} \quad \mathscr{D}_{\text{stadium}} = \text{rectangle} + \text{semicircular ends}\,,
\end{eqnarray*}
creates a smooth, convex, but non-integrable boundary which is the source of classical chaos in the system. The boundary condition enforces Dirichlet walls, and the vector potential is taken in the symmetric gauge. When a transverse magnetic field is applied, the Hamiltonian is given by \eqref{mag-ham}. The system is diagonalized numerically in a truncated basis appropriate for the stadium geometry and magnetic field strength. Using the resulting eigenstates and eigenvalues, the matrix elements of the canonical position and momentum operators $x$ and $p$ are computed, and the full Heisenberg evolution $x(t)$ is evaluated explicitly. The thermal OTOC is computed via the algorithm above and for all field strengths, we use a consistent thermal cutoff to ensure fair comparisons across spectra of differing bandwidths. Our results for $C_{T}(t)$ are plotted for a range of magnetic field strengths and temperatures in Fig. \ref{figure:magnetic-stadium} from which we note that:\\

\begin{itemize}
    \item At $B=0$ (Fig. \ref{fig:stadium1}), the stadium billiard is a classic chaotic system. The OTOC exhibits clear initial growth followed by plateauing or saturation. This is consistent with the Lyapunov-like early-time growth expected in chaotic systems and eventual saturation due to a finite Hilbert space size and energy truncation. Since there is no magnetic field, both bulk and edge dynamics are entangled in the OTOC.
    \item As the magnetic field strength is increased, the nature of operator growth encoded in the thermal OTOC changes qualitatively. At low (but non‑zero) $B$, the OTOC $C_T(t)$ exhibits rapid early‑time growth consistent with chaotic operator spreading, closely resembling the field‑free stadium case. In the intermediate regime $B \sim 1.5\text{–}3.0$, early‑time growth persists but is followed by increasing curvature and the emergence of structured, quasi‑periodic fluctuations rather than a true saturation plateau. At large fields $B \gtrsim 3.5$, the OTOC is dominated by pronounced oscillations with characteristic timescales set by the inverse cyclotron frequency $\omega_c^{-1}$, signaling a crossover from chaotic to coherent, magnetically constrained dynamics.\\

    \noindent
    This behavior can be understood from the evolution of the energy spectrum with increasing magnetic field. As $B$ grows, the spectrum transitions from irregular, stadium‑dominated level spacings to Landau‑like bands with approximately harmonic spacing,
    \begin{eqnarray}
        E_n \sim \hbar \omega_c \left( n + \tfrac{1}{2} \right)\,,
    \end{eqnarray}
    with $\omega_c = \frac{eB}{m}$. Such a quasi‑harmonic structure gives rise to coherent interference effects in time‑dependent observables. In particular, for operators such as $x(t)$, the spectral decomposition involves sums over matrix elements weighted by phases $e^{i(E_\alpha - E_\beta)t/\hbar}$, which constructively interfere at timescales related to $2\pi/\omega_c$. As a result, instead of relaxing to a constant value, the OTOC exhibits long‑lived oscillations whose mean value increases with $B$. This reflects not enhanced scrambling, but the growing influence of magnetic rigidity and intrinsic operator non‑commutativity in the Landau‑dominated regime. In this limit, cyclotron motion constrains phase‑space exploration, suppressing chaotic spreading while reorganizing operator growth into a regular, oscillatory pattern rather than exponential behavior.
    \item At low temperatures ($\beta \gg 1$) on the other hand, the thermal ensemble is dominated by the system’s low-lying eigenstates, which are typically delocalized over the entire billiard (see Appendix §\ref{app:num3}); symmetric and non-chaotic, reflecting remnants of integrable motion and insensitive to fine boundary structure, since their wavelengths are large compared to the boundary curvature scale. As a result, the OTOC in this regime tends to grow slowly, exhibiting sublinear or diffusive-like behavior in time. This reflects the absence of strong mode mixing or operator growth, consistent with the regular structure of the contributing eigenstates.
    \item As the temperature increases, more excited states contribute to the thermal average. These higher-energy states exhibit
	greater angular momentum and nodal  structure, with amplitudes concentrated near the stadium boundary. They are also sensitive to the exact geometry, including curvature of the caps and the discontinuity between circular and straight segments. In this regime, the OTOC grows more rapidly with a clear exponential regime at early times. 
    \item At yet higher temperatures (small $\beta$), even more localized and rapidly varying states contribute. However, in the presence of a strong transverse magnetic field $B$, the particle dynamics experience increasing magnetic rigidity resulting in the kinetic motion being suppressed perpendicular to the magnetic field. In this regime, cyclotron orbits dominate, with a suppression of transverse diffusion.
    Consequently, even as more states are thermally populated, the onset of Landau quantization and noncommutative geometry leads to a suppression of scrambling, and the OTOC growth is again damped. The Lyapunov exponent $\lambda_L$ is expected to exhibit a non-monotonic dependence on temperature and magnetic field, increasing with $T$ before being suppressed by $B$.
\end{itemize}

\subsection{Temperature and $B$-field dependence of the OTOC growth}
We extracted the OTOC growth exponent\footnote{A word of caution: while the early-time growth of the thermal OTOC in the magnetic stadium appears visually exponential for moderate values of $B$, especially at elevated temperatures, such growth is numerically subtle and can often be mimicked by power laws over short time windows. This echoes the findings of Hashimoto et al. for $B = 0$, where the OTOC scales as $t^2$ rather than $e^{\lambda t}$. In the absence of many-body interactions or additional semiclassical limits, the extracted Lyapunov-like exponents should thus be interpreted as effective short-time indicators of scrambling, rather than genuine Lyapunov growth. It is in this sense that the exponent we refer to in this section should be understood.} $\lambda_L(T, B)$ from the early-time growth of the thermal OTOC $C_T(t) = -\langle [x(t), p]^2 \rangle_\beta$, evaluated numerically for the charged particle in a stadium billiard subject to a transverse magnetic field as above. As discussed in the Introduction, this quantity serves as a diagnostic of operator growth and quantum scrambling in the underlying dynamics. In our context, the system is a single charged particle confined to a stadium-shaped billiard, a paradigmatic example of a classically chaotic system, but in the presence of a perpendicular magnetic field. The magnetic field modifies the dynamics via the Lorentz force, introducing cyclotron motion and thereby competing with the chaotic effects of the stadium boundary. At early times, the OTOC typically exhibits a regime of exponential growth where $C_T(t) \sim A e^{2\lambda_L t} + C_0$, before saturating due to finite Hilbert space effects or thermal suppression. We extract $\lambda_L$ by performing a log-linear fit to $\log[C_T(t) - C_0]$ over a suitable time window $[t_{\text{min}}, t_{\text{max}}]$, chosen to lie within the clean exponential growth regime.
$t_{\text{min}}$ is taken after initial transients, and $t_{\text{max}}$ is set before saturation begins. The fit is tested for stability under small changes in this window.\\

\begin{figure}
\centering
\includegraphics[width=0.5\textwidth]{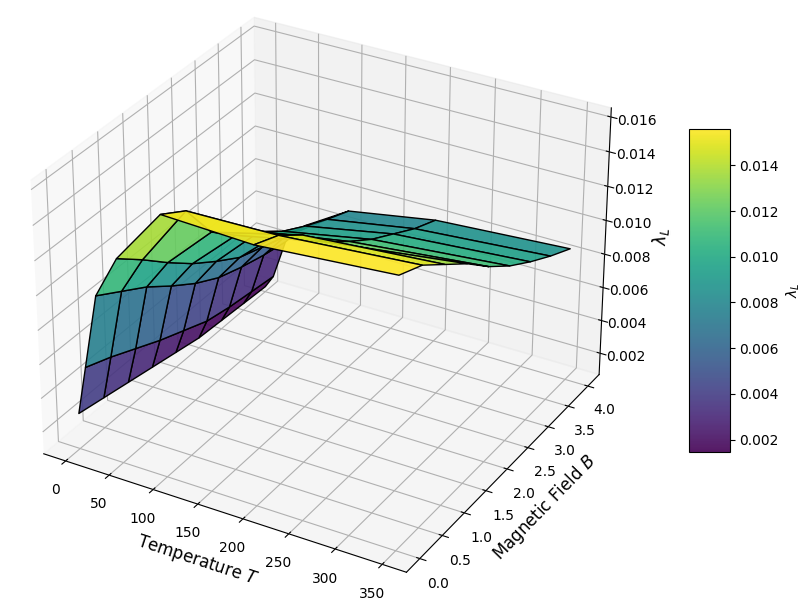}
\caption{Thermal growth exponent $\lambda_L(T, B)$ extracted from the early-time growth of the thermal OTOC $C_T(t) = -\langle [x(t), p]^2 \rangle_\beta$ for a charged particle in a stadium billiard with transverse magnetic field. The surface plot shows how the extracted exponent varies as a function of temperature $T$ and magnetic field strength $B$.}
\label{Lyapunov-surface}
\end{figure}

\noindent
The resulting exponents $\lambda_L(B, T)$ are compiled into the surface plot in Fig. \ref{Lyapunov-surface} and reveals revealing rich non-monotonic structure. At low temperatures (large $\beta$), only the lowest-lying eigenstates contribute significantly to the thermal trace. These states are typically delocalized and less sensitive to the boundary-induced chaos of the stadium geometry. As a result, the extracted $\lambda_L$ remains small, reflecting the predominantly regular dynamics in this low-energy sector. As temperature increases, higher-energy eigenstates—more strongly influenced by the chaotic boundary geometry—enter the ensemble. The thermal OTOC exhibits faster early-time growth, and $\lambda_L$ increases correspondingly. This growth reflects a classical-to-quantum correspondence where classical trajectories in the stadium are known to be chaotic, and the quantum analog inherits this behaviour through eigenstate structure and spectral correlations.\\

\noindent
At each fixed temperature however, increasing the magnetic field strength $B$ systematically suppresses the growth exponent. This suppression is a consequence of magnetic rigidity in which the transverse Lorentz force increasingly confines trajectories to cyclotron orbits, thereby limiting the phase-space exploration necessary for efficient scrambling. For large $B$, $\lambda_L$ decays significantly and eventually saturates near zero, signalling the restoration of regular dynamics. The joint surface plot of $\lambda_L(T, B)$ in Fig. \ref{Lyapunov-surface} reveals this interplay between thermal activation and magnetic suppression. At fixed $T$, $\lambda_L$ decreases monotonically with $B$, while at fixed $B$, it increases and eventually saturates with $T$. The surface is consistent with a smooth thermal-to-magnetic crossover separating a chaotic high-temperature regime from a magnetically regularized phase. We do not observe any evidence of a sharp transition or instability.

\section{Guiding-center OTOCs in a transverse magnetic field}
\label{sec:guiding-center}
Out-of-time-ordered correlators are powerful probes of scrambling and  chaos in quantum mechanics. As we have seen above, they can be formulated exclusively in terms of position and momentum operators making them amenable to efficient numerical computation. In the presence of a strong transverse magnetic field, however, the natural degrees of freedom are reorganized into \textit{guiding center coordinates}, non-commuting operators that encode the center of cyclotron motion in the lowest Landau level \cite{PhysRevLett.100.246802}. These operators furnish a uniquely geometric language for studying quantum dynamics in magnetic systems, with a built-in separation between bulk Landau-level physics and boundary or potential-induced structure.

\subsection{Guiding-center coordinates}
In this section, we introduce and compute OTOCs constructed from guiding center operators, focusing on their ability to capture nontrivial scrambling dynamics purely induced by the underlying non-commutative geometry. These correlators are sensitive to both the intrinsic curvature of the Landau orbits and to perturbations, such as boundaries, confining potentials and disorder, that couple distinct Landau levels. Crucially, this formulation offers a clean diagnostic of where and how scrambling occurs whether in the bulk, where dynamics are highly constrained by magnetic translation symmetry, or near edges, where boundary conditions and curvature of the classical orbits introduce rich dynamical structure.\\ 

\noindent
Toward this end, consider again a charged particle in a uniform magnetic field with Hamiltonian \eqref{mag-ham}. It is convenient to introduce the mechanical (kinetic) momentum
$\bm{\Pi} = \bm p - q \bm A$,
whose components satisfy the non‑trivial commutation relation
\begin{eqnarray}
    [\Pi_x,\Pi_y] = i\hbar qB = i\frac{\hbar^2}{\ell_B^2}\,,
\end{eqnarray}
with magnetic length $\ell_B = \sqrt{\frac{\hbar}{|q|B}}$. The guiding center coordinates $\bm R=(X,Y)$, representing the center of cyclotron motion, are defined as
\begin{eqnarray}
    \bm R = \bm r - \frac{\ell_B^2}{\hbar}\,\hat z \times \bm\Pi,
\end{eqnarray}
or explicitly,
\begin{eqnarray}
    X = x - \frac{\ell_B^2}{\hbar}\Pi_y\,,
    \qquad
    Y = y + \frac{\ell_B^2}{\hbar}\Pi_x\,.
\end{eqnarray}
It is straightforward to check that the guiding center coordinates commute with the kinetic momentum, $[X,\Pi_x]=[X,\Pi_y]=[Y,\Pi_x]=[Y,\Pi_y]=0$, and satisfy the intrinsic non‑commutative algebra
\begin{eqnarray}
    [X,Y] = -i\ell_B^2\,.
    \label{non-comm-alg}
\end{eqnarray}
This non‑commutativity reflects the fundamental quantum geometry induced by the magnetic field and implies a minimal area $\ell_B^2$ in the guiding‑center phase space. For the free particle, $V=0$, the Hamiltonian depends only on $\boldsymbol{\Pi}$, as $H_0 = \bm{\Pi}^2/2m$,
and therefore $[H_0,X]=[H_0,Y]=0$. The guiding centers are constants of motion in the bulk, and any non‑trivial dynamics must arise from external potentials or boundary effects.\\

\noindent
To probe scrambling associated with the guiding‑center degrees of freedom, we consider the out‑of‑time‑ordered correlator
$C_R(t) = -\left\langle [X(t),Y(0)]^2 \right\rangle$, where operators evolve in the Heisenberg picture and the expectation value may be taken in a microcanonical, thermal, or general stationary state. This OTOC directly probes the time evolution of the non‑commutative geometry encoded in $[X,Y]$, rather than the kinetic cyclotron motion. The Heisenberg evolution of $X(t)$ is given as usual by
$X(t) = e^{iHt/\hbar} X e^{-iHt/\hbar}$.
Since $X$ commutes with the kinetic Hamiltonian, any time dependence arises entirely from the potential $V(\bm r)$. Using
$X = x - \frac{\ell_B^2}{\hbar}\Pi_y$, we find that
\begin{eqnarray}
    [V(\bm r),X] = -\frac{\ell_B^2}{\hbar}[V(\bm r),\Pi_y]\,,
\end{eqnarray}
and because $V(\bm r)$ depends only on position,
$[V(\mathbf r),\Pi_y] = [V,p_y] = i\hbar\,\partial_y V$,
and hence $[H,X] = -i\ell_B^2\,\partial_y V$.
Similarly, $[H,Y] = i\ell_B^2\,\partial_x V$.
These equations show that the guiding center motion is driven entirely by gradients of the potential, $\dot{\bm R} \propto \hat z \times \nabla V$, which is the quantum analogue of the classical $\bm E \times \bm B$ drift.\\

\noindent
To extract the early‑time behavior of the OTOC, we expand $X(t)$ using the BCH formula,
\begin{eqnarray}
    X(t) = X + \frac{it}{\hbar}[H,X]
+ \frac{(it)^2}{2!\hbar^2}[H,[H,X]] + \cdots.
\end{eqnarray}
The first nested commutator is simply $[H,X] = -i\ell_B^2\,\partial_y V$. The second nested commutator involves $[H,\partial_y V]$,
which gives terms proportional to second derivatives of the potential and the kinetic momentum. Explicitly,
\begin{eqnarray}
    [H,\partial_y V]
    = -\frac{i\hbar}{2m}
    \left(
    \Pi_x\,\partial_x\partial_y V
    + \partial_x\partial_y V\,\Pi_x
    + \Pi_y\,\partial_y^2 V
    + \partial_y^2 V\,\Pi_y
    \right)\,.
\end{eqnarray}
Consequently,
\begin{eqnarray}
    [H,[H,X]]
    = -\frac{\hbar\ell_B^2}{2m}
    \left(
    \Pi_x\,\partial_x\partial_y V
    + \partial_x\partial_y V\,\Pi_x
    + \Pi_y\,\partial_y^2 V
    + \partial_y^2 V\,\Pi_y
    \right).
\end{eqnarray}
We now compute the commutator entering the OTOC,
\begin{eqnarray}
    [X(t),Y]
    = [X,Y]
    + \frac{it}{\hbar}[[H,X],Y]
    + \frac{(it)^2}{2!\hbar^2}[[H,[H,X]],Y]
    + \cdots.
\end{eqnarray}
The zeroth‑order term is the constant
$[X,Y] = -i\ell_B^2$. The first‑order correction is $[[H,X],Y]
= [-i\ell_B^2\,\partial_y V,Y]
= -i\ell_B^2[\partial_y V,Y]$.
Then, using
$[\partial_y V,Y] = -i\ell_B^2\,\partial_x\partial_y V$, and $[\partial_y V,Y] = -i\ell_B^2\,\partial_x\partial_y V$, we obtain
\begin{eqnarray}
    [[H,X],Y] = -\ell_B^4\,\partial_x\partial_y V\,.
\end{eqnarray}
Keeping terms up to linear order in $t$, the commutator becomes
\begin{eqnarray}
    [X(t),Y]
\approx
-i\ell_B^2
\left(
1 + \frac{\ell_B^2}{\hbar} t\,\partial_x\partial_y V
\right)
+ \mathcal O(t^2)\,.
\end{eqnarray}
Squaring this expression and taking the expectation value yields the early‑time OTOC,
\begin{eqnarray}
    C_R(t)
    = -\langle [X(t),Y]^2\rangle
    \approx
    \ell_B^4
    \left(
    1 + \frac{2\ell_B^2}{\hbar} t\,\langle \partial_x\partial_y V\rangle
    + \mathcal O(t^2)
    \right)\,.
    \label{guigeOTOC}
\end{eqnarray}
Several important conclusions follow immediately from \eqref{guigeOTOC}. First, for the free particle ($V=0$) or in the bulk of a Landau level, the guiding‑center commutator is time‑independent,
\begin{eqnarray}
    [X(t),Y]=[X,Y]\quad
    \Rightarrow\quad
    C_R(t)=\ell_B^4\,.
\end{eqnarray}
There is no scrambling because the non‑commutative geometry is a static property of the Hilbert space. Second, time dependence, and consequently scrambling, arises only through external potentials or boundary confinement. Disorder, smooth confining potentials, or hard‑wall billiard boundaries induce guiding‑center drift and lead to non‑trivial OTOCs. Third, the early‑time growth of the guiding‑center OTOC is linear in time, not quadratic, with a coefficient controlled by the  second derivative $\partial_x\partial_y V$. Physically, this quantity measures the shear of the potential landscape, which governs the sensitivity of guiding‑center trajectories.
Finally, the strong dependence on $\ell_B^6$ implies that guiding‑center scrambling is also suppressed at large magnetic fields. As the cyclotron radius shrinks, drift becomes slower and the guiding‑center dynamics become increasingly rigid.

\subsection{Guiding-Center OTOCs and bulk–edge separation}
\label{sec:edge}
The OTOCs in \cite{Hashimoto:2017oit} were constructed from position and momentum operators, such as $[x(t), p]^2$, and interpreted as diagnostics of sensitivity to initial conditions in quantum systems. However, in the presence of a strong magnetic field, these operators no longer commute in a gauge-invariant way, and their dynamics inevitably mix the fast cyclotron motion with slower, large-scale drift. Moreover, position-space OTOCs do not distinguish between the dynamics occurring deep in the bulk and the rich edge behavior introduced by confinement or geometry. This limits their utility in systems such as quantum Hall fluids or magnetic billiards, where boundaries and non-commutative geometry play a key role.\\

\noindent
Guiding-center OTOCs offer a conceptually cleaner alternative with several key advantages. First, they are manifestly basis-independent and gauge-invariant, since they are constructed entirely from commutators of the Hamiltonian and coordinate operators. Second, guiding centers commute with the kinetic momentum operators $\Pi_i$, effectively decoupling them from the fast cyclotron dynamics and isolating the slow drift of orbital motion which is exactly the physics that governs edge dynamics and long-time transport. Third, the guiding-center commutator $[X, Y] = -i\ell_B^2$ defines a natural scale $\ell_B^4$ for the OTOC, making deviations from this value physically meaningful and easy to interpret. In the bulk of a clean system, where the potential is flat or absent, the guiding-center OTOC remains constant, $C_R(t) = \ell_B^4$. This reflects the rigidity of the non-commutative geometry and the absence of information scrambling since operator growth is essentially frozen. However, once boundaries or inhomogeneous potentials are introduced, the guiding-center coordinates acquire time dependence, as can be seen in \eqref{guigeOTOC}. The early-time behavior is particularly telling. The OTOC grows linearly in time, with a rate proportional to the mixed second derivative $\partial_x \partial_y V$ of the potential,
\begin{eqnarray}
    \frac{d}{dt}C_R(t) \propto \ell_B^6 \langle \partial_x\partial_y V \rangle\,.
\end{eqnarray}
This derivative encodes the local shear of the potential landscape and governs how rapidly nearby guiding-center trajectories diverge. In this sense, the guiding-center OTOC is a sensitive probe of geometric deformation and confinement.\\

\noindent
This intrinsic link to curvature and shear makes guiding-center OTOCs ideally suited for a range of applications. In confined geometries, they directly track the onset of edge dynamics and skipping orbits. In quantum billiards, they diagnose the sensitivity of edge modes to boundary curvature and chaotic scattering. And in disordered systems, they provide a window into how disorder competes with topological protection to induce or suppress scrambling. In each case, the guiding-center OTOC cleanly separates the dynamical effects of boundaries and potentials from the topologically inert bulk, revealing the structure of quantum information flow in magnetic systems with  clarity.\\

\noindent
To illustrate, we evaluate the guiding-center OTOC $C_R(t)$ for individual eigenstates of the, by now familiar, charged particle confined to a circular billiard of radius $R$, in the presence of a perpendicular magnetic field $B$. As we have seen already, this geometry is analytically tractable and, in the case of the guiding-center OTOC, permits clean identification of bulk and edge states. Recall that the eigenstates of the Hamiltonian \eqref{symmetric-gauge-ham} are labeled by radial and angular quantum numbers $(n_r, m)$, and take the form given in \eqref{eq:bulkpsi} which arise as the most general solutions of the radial Schrödinger equation in a uniform magnetic field. Imposing normalizability and the hard-wall boundary condition at the billiard radius quantizes the parameter a to a non‑negative integer, $a=n_r$. In this case the confluent hypergeometric function reduces to a generalized Laguerre polynomial via the identity, 
\begin{eqnarray}
    {}_1F_1(-n; \alpha + 1; z) = \frac{n!}{(\alpha+1)_n} L_n^\alpha(z)
\quad\text{for } n\in\mathbb{N}_0\,,
\end{eqnarray}
where $(\alpha+1)_n$ is the Pochhammer symbol. Consequently, the wavefunctions take the equivalent Landau‑level–adapted form
\begin{eqnarray}
    \psi_{n_r, m}(r, \theta) = \mathcal{N}_{n_r, m} \, r^{|m|}\, e^{im\theta}\, e^{-r^2/(4\ell_B^2)} \,L_{n_r}^{|m|}\!\left(\frac{r^2}{2\ell_B^2}\right)\,,
    \label{LL-wavefn}
\end{eqnarray}
with radial probability density
\begin{eqnarray}
    |\psi_{n_r,m}(r)|^2 \propto r^{2|m|} e^{-r^2/(2\ell_B^2)} \left[ L_{n_r}^{|m|} \left( \frac{r^2}{2\ell_B^2} \right) \right]^2\,.
\end{eqnarray}
In both cases, $r\in [0,R]$ and Dirichlet boundary conditions $\psi(R) = 0$ discretize the allowed states. While both the confluent hypergeometric form and the associated Laguerre form describe eigenstates of a charged particle in a circular magnetic billiard, they differ in emphasis. The former arises naturally from solving the Schrödinger equation with magnetic field and hard-wall boundary conditions, while the latter organizes states according to Landau level structure, with radial quantum number $n_r$ counting Landau level excitations and $|m|$ capturing angular localization. For guiding-center OTOCs, the Laguerre basis is preferable because it cleanly separates cyclotron dynamics from guiding-center motion, making it ideally suited to isolating bulk–edge physics.\\

\begin{figure}
\centering
\includegraphics[width=0.5\textwidth]{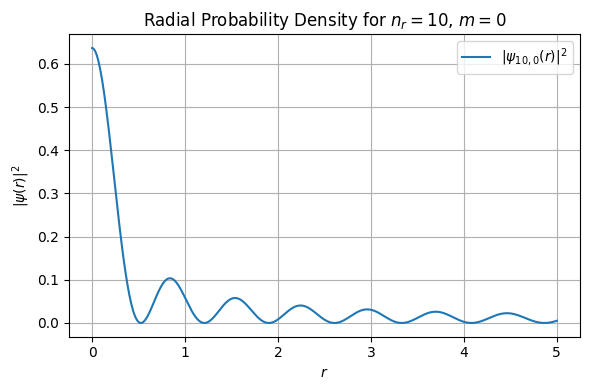}
\caption{Radial probability density $|\psi_{10,0}(r)|^2$ for the circular magnetic billiard state with $n_r = 10, m = 0$.}
\label{BulkRPD}
\end{figure}
\begin{figure}
\centering
\includegraphics[width=0.5\textwidth]{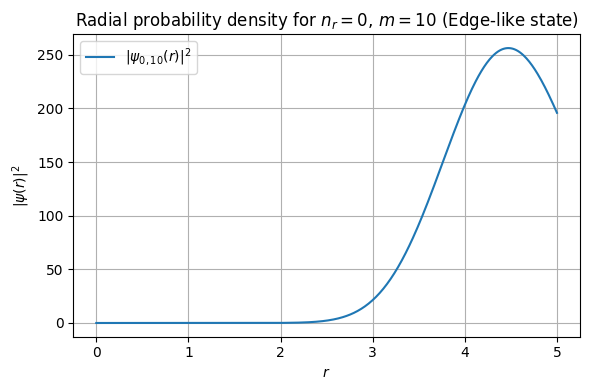}
\caption{Radial probability density $|\psi_{0,10}(r)|^2$ for the edge-like state with $n_r = 0, m = 10$ strongly peaked near the boundary.}
\label{EdgeRPD}
\end{figure}

\begin{figure}
\centering
\includegraphics[width=0.6\textwidth]{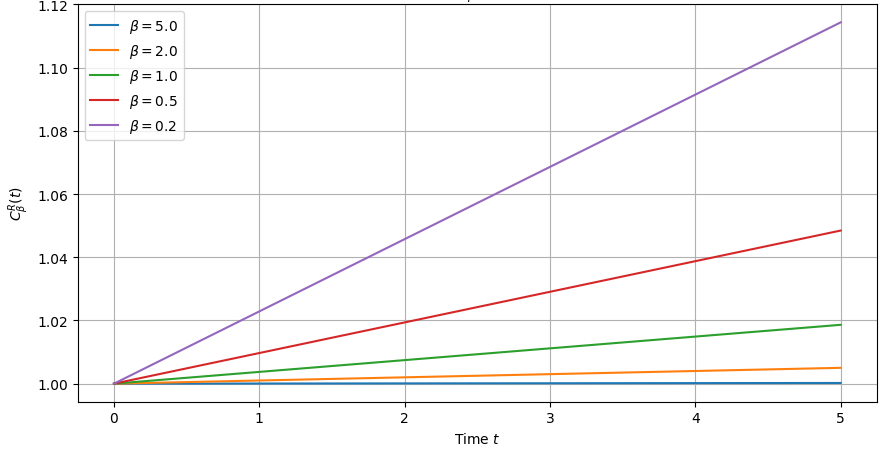}
\caption{Thermal guiding-center out-of-time-order correlator (OTOC) $C^{R}_{\beta}(t)$ for a charged particle confined in a circular magnetic billiard with hard-wall boundary conditions. The guiding-center OTOC is computed from the thermal expectation of the squared commutator between the guiding-center coordinates $X(t)$ and $Y(0)$, using the magnetic algebra $[X, Y] = i \ell_B^2$. The linear-in-time growth reflects the effective semiclassical dynamics of guiding centers in a nonuniform potential $V(x, y)$, where the curvature $\langle \partial_x \partial_y V \rangle_\beta$ is computed explicitly in the energy eigenbasis. The slope increases with temperature, indicating enhanced thermal mixing of high-energy eigenstates localized near the hard wall.}
\label{Thermal-OTOC}
\end{figure}

\noindent
When $|m|$ is small in \eqref{LL-wavefn}, the prefactor $r^{2|m|}$ does not strongly suppress amplitude near the origin. When $n_r$ is large, the Laguerre polynomial introduces many oscillations and the wavefunction is spread out, but still centered closer to the interior of the disk. These states are analogous to bulk Landau level wavefunctions and are localized around finite radii well within the boundary (see Fig. \ref{BulkRPD}). These are \textit{bulk states}. On the other hand, large $|m|$ strongly suppresses the amplitude near the origin via the prefactor $r^{2|m|}$. The Gaussian tail however forces localization near a peak at $r_{\text{peak}} \sim \sqrt{2|m|}\, \ell_B$. For a fixed system size $R$, states with $r_{\text{peak}} \approx R$ are peaked near the boundary\footnote{One might expect that Dirichlet boundary conditions should force wavefunctions to vanish at the physical boundary of the billiard. However, this expectation applies to the full wavefunction, not directly to the guiding-center coordinate probability since the guiding center is a projected coordinate, defined in the presence of a strong magnetic field. So even if the full wavefunction vanishes at the hard wall boundary, the guiding-center component of the probability density does not have to.} (as in Fig. \ref{EdgeRPD}). These are \textit{edge states}.  Importantly, the hard-wall confinement lifts the degeneracy of Landau levels and states localized near the boundary generally have higher energy than those in the bulk. As a result, at low temperatures (large $\beta$) we expect that the thermal ensemble is dominated by low-energy bulk-like states, while edge-like states contribute significantly only at higher temperatures when thermal excitation allows occupation of higher-energy levels. This spectral ordering ensures that the guiding-center OTOC, $C^R_\beta(t)$, exhibits qualitatively distinct behavior in the low- and high-temperature regimes, reflecting the underlying spatial structure of the eigenstates.
These results are captured in Fig. \ref{Thermal-OTOC}.

\section{Conclusions}
\noindent
While the formalism for computing OTOCs in single-particle quantum mechanics was laid out by Hashimoto \textit{et al.}~\cite{Hashimoto:2017oit}, their analysis revealed that even for classically chaotic systems such as the stadium billiard, the thermal OTOC $C_T(t) = -\langle [x(t), p]^2 \rangle_\beta$ saturates rather than grows exponentially. In this work, we build on that foundation by incorporating a transverse magnetic field and systematically exploring the $(T,B)$ parameter space. We find that the presence of the magnetic field introduces a tunable rigidity that modulates the chaotic dynamics and, crucially, uncovers an effective early-time exponential growth regime in $C_T(t)$ at finite temperature. This allows us to extract Lyapunov-like exponents $\lambda_L(T,B)$, and construct a surface that captures the thermal and magnetic control of quantum scrambling. In this way, our study extends the conclusions of~\cite{Hashimoto:2017oit}, demonstrating that chaos in quantum billiards, though absent in strict saturation-based diagnostics, can emerge in a controlled sense when additional physical ingredients are included.\\

\noindent
Specifically, at low magnetic field strengths, the canonical OTOC displays clear growth indicative of quantum scrambling in a chaotic system. As the magnetic field increases, this growth is suppressed, and $\lambda_L$ diminishes, consistent with the emergence of cyclotron motion and the suppression of boundary-induced chaos due to magnetic rigidity. This suppression is most pronounced at high temperatures, where edge-localized and disorder-sensitive eigenstates dominate the thermal ensemble. In contrast, low-temperature dynamics are governed by delocalized, low-lying states, resulting in smoother OTOC profiles and reduced sensitivity to boundary curvature. The full $\lambda_L(T, B)$ surface offers a rare, controlled view into the field–temperature interplay in quantum chaotic systems.\\

\noindent
In parallel, we investigated a structurally distinct OTOC constructed from the guiding-center coordinates $(X, Y)$ that emerge upon Landau-level projection. For the circular magnetic billiard, we derived and evaluated an explicit expression for $C^R_\beta(t) = -\langle [X(t), Y]^2 \rangle_\beta$, which is closely related to the mixed derivatives of the confining potential $V(r)$. The resulting OTOC displays only linear or sub-linear growth with time and exhibits no evidence of a Lyapunov regime. This provides a striking contrast to the canonical case and highlights the stabilizing role of the guiding center under strong magnetic field. The comparison thus reveals how chaos is not only suppressed by magnetic rigidity but also filtered out entirely in observables that are naturally defined within the lowest Landau level.\\ 

\noindent
More broadly, our results emphasize that OTOCs are not universal diagnostics of chaos but rather probe operator-specific aspects of dynamical growth. The stark contrast between the canonical OTOC, constructed from position and momentum operators $(x, p)$, and the guiding-center OTOC, built from projected coordinates $(X, Y)$, illustrates this point concretely. While the former exhibits exponential growth at early times, with a well-defined thermal growth exponent $\lambda_L$ that is sensitive to both geometry and field strength, the latter shows no such exponential regime, even in chaotic settings. This is not merely a feature of the underlying dynamics, but a reflection of what degrees of freedom are being probed. The guiding-center operators are tailored to Landau-level-projected physics and effectively integrate out fast cyclotron motion \cite{PhysRevLett.100.246802,PhysRevB.49.16765}, filtering out short-time chaotic sensitivity. Conversely, canonical operators retain sensitivity to full phase-space evolution, including boundary-induced scrambling. Thus, different operator choices encode different dynamical structures \cite{Sabella-Garnier:2019tsi,Bukva:2019oly}, and care must be taken in interpreting OTOCs. Indeed, they are not absolute measures of chaos but are best understood as context-dependent probes of operator growth in specific kinematic sectors.\\

\noindent
Beyond offering a new diagnostic lens on quantum scrambling in billiard systems, our results point to an interesting interplay between quantum chaos and magnetic rigidity. The exponent $\lambda_L$, extracted from the early-time growth of the canonical OTOC $C_T(t)$, reveals a smooth crossover from boundary-induced chaotic dynamics at low magnetic field to a magnetically dominated regime where chaos is progressively suppressed. This transition is tunable, coherent, and rooted in the emergent cyclotron motion that restricts phase-space exploration. Viewed this way, the transverse magnetic field acts as a controllable dial on quantum chaoticity, hinting at practical routes to engineer or suppress chaos in mesoscopic or quantum technological platforms. This raises intriguing possibilities for quantum control, including stabilizing scrambling in quantum simulators or even manipulating dynamical sensitivity in information-processing tasks. Future work in this direction could explore this magnetic tuning in Floquet-driven systems, connections to transport and thermalization \cite{PhysRevA.97.042330}, or even exploit it to identify emergent integrable structures in the high-field limit. Our framework here opens a controlled and computationally accessible arena to investigate the controllability of chaos in quantum systems.

\acknowledgments
We would like to thank Nitin Gupta and Jaco Van Zyl for useful discussions at various points. JM is supported in part by the “Quantum Technologies for Sustainable Development” grant
from the National Institute for Theoretical and Computational Sciences of South Africa
(NITHECS). CB is supported by the Oppenheimer Memorial Trust, and Harry Crossley Research Fellowships. JM and CB would like to acknowledge support from the ICTP, Trieste through the Associates Programme and from the Simons Foundation through grant number 284558FY19. Computations were performed using facilities provided by the University of Cape Town’s ICTS High Performance Computing team: hpc.uct.ac.za  -- \url{https://doi.org/10.5281/zenodo.10021612}


 \bibliographystyle{JHEP}
\bibliography{biblio.bib}

\providecommand{\href}[2]{#2}\begingroup\raggedright\begin{thebibliography}{10}

\bibitem{Hashimoto:2017oit}
K.~Hashimoto, K.~Murata and R.~Yoshii, \emph{{Out-of-time-order correlators in
  quantum mechanics}},
  \href{https://doi.org/10.1007/JHEP10(2017)138}{\emph{JHEP} {\bfseries 10}
  (2017) 138} [\href{https://arxiv.org/abs/1703.09435}{{\ttfamily
  1703.09435}}].

\bibitem{Berry:1977pr}
M.~Berry, \emph{Regular and irregular motion},
  \href{https://doi.org/10.1063/1.523334}{\emph{AIP Conference Proceedings}
  {\bfseries 46} (1977) 16}.

\bibitem{Gutzwiller:1990ch}
M.C.~Gutzwiller, \emph{Chaos in Classical and Quantum Mechanics},
  Springer-Verlag (1990).

\bibitem{Zaslavsky:1981eh}
G.M.~Zaslavsky, \emph{Stochasticity in quantum systems},
  \href{https://doi.org/10.1016/0370-1573(81)90121-8}{\emph{Physics Reports}
  {\bfseries 80} (1981) 157}.

\bibitem{Shepelyansky:1994hy}
D.L.~Shepelyansky, \emph{Quantum chaos and localization}, {\emph{Physica D}
  {\bfseries 83} (1994) 208}.

\bibitem{Haake:2001qp}
F.~Haake, \emph{Quantum Signatures of Chaos}, Springer (2001).

\bibitem{Mehta:2004sp}
M.L.~Mehta, \emph{Random Matrices}, Elsevier (2004).

\bibitem{CASATI1999293}
G.~Casati and T.~Prosen, \emph{The quantum mechanics of chaotic billiards},
  \href{https://doi.org/https://doi.org/10.1016/S0167-2789(99)00002-0}{\emph{Physica
  D: Nonlinear Phenomena} {\bfseries 131} (1999) 293}.

\bibitem{Heller:1984ws}
E.J.~Heller, \emph{Bound-state eigenfunctions of classically chaotic
  hamiltonian systems: Scars of periodic orbits},
  \href{https://doi.org/10.1103/PhysRevLett.53.1515}{\emph{Phys. Rev. Lett.}
  {\bfseries 53} (1984) 1515}.

\bibitem{Chirikov:1981qm}
B.~Chirikov, \emph{A universal instability of many-dimensional oscillator
  systems}, {\emph{Physics Reports} {\bfseries 52} (1981) 263}.

\bibitem{Lima_2005}
E.F.d.~Lima and J.E.M.~Hornos, \emph{Matrix elements for the morse potential
  under an external field},
  \href{https://doi.org/10.1088/0953-4075/38/7/004}{\emph{Journal of Physics B:
  Atomic, Molecular and Optical Physics} {\bfseries 38} (2005) 815}.

\bibitem{PhysRevE.53.4555}
M.A.M.~de~Aguiar, \emph{Eigenvalues and eigenfunctions of billiards in a
  constant magnetic field},
  \href{https://doi.org/10.1103/PhysRevE.53.4555}{\emph{Phys. Rev. E}
  {\bfseries 53} (1996) 4555}.

\bibitem{PhysRevLett.100.246802}
B.A.~Bernevig and F.D.M.~Haldane, \emph{Model fractional quantum hall states
  and jack polynomials},
  \href{https://doi.org/10.1103/PhysRevLett.100.246802}{\emph{Phys. Rev. Lett.}
  {\bfseries 100} (2008) 246802}.

\bibitem{PhysRevB.49.16765}
J.~Yang, \emph{Wave functions of hierarchy states in the fractional quantum
  hall effect}, \href{https://doi.org/10.1103/PhysRevB.49.16765}{\emph{Phys.
  Rev. B} {\bfseries 49} (1994) 16765}.

\bibitem{Sabella-Garnier:2019tsi}
P.~Sabella-Garnier, K.~Schalm, T.~Vakhtel and J.~Zaanen,
  \emph{{Thermalization/Relaxation in integrable and free field theories: an
  Operator Thermalization Hypothesis}},
  \href{https://arxiv.org/abs/1906.02597}{{\ttfamily 1906.02597}}.

\bibitem{Bukva:2019oly}
A.~Bukva, P.~Sabella-Garnier and K.~Schalm, \emph{{Operator thermalization vs
  eigenstate thermalization}},
  \href{https://arxiv.org/abs/1911.06292}{{\ttfamily 1911.06292}}.

\bibitem{PhysRevA.97.042330}
E.~Iyoda and T.~Sagawa, \emph{Scrambling of quantum information in quantum
  many-body systems},
  \href{https://doi.org/10.1103/PhysRevA.97.042330}{\emph{Phys. Rev. A}
  {\bfseries 97} (2018) 042330}.

\bibitem{baratta2023dolfinx}
I.A.~Baratta, J.P.~Dean, J.S.~Dokken, M.~Habera, J.S.~Hale, C.N.~Richardson
  et~al., \emph{{DOLFINx}: The next generation {FEniCS} problem solving
  environment}, \href{https://doi.org/10.5281/zenodo.10447666}{\emph{Preprint}
  (2023) }.

\bibitem{scroggs2022basix}
M.W.~Scroggs, I.A.~Baratta, C.N.~Richardson and G.N.~Wells, \emph{{Basix}: a
  runtime finite element basis evaluation library},
  \href{https://doi.org/10.21105/joss.03982}{\emph{Journal of Open Source
  Software} {\bfseries 7} (2022) 3982}.

\bibitem{scroggs2022construction}
M.W.~Scroggs, J.S.~Dokken, C.N.~Richardson and G.N.~Wells, \emph{Construction
  of arbitrary order finite element degree-of-freedom maps on polygonal and
  polyhedral cell meshes}, \href{https://doi.org/10.1145/3524456}{\emph{ACM
  Transactions on Mathematical Software} {\bfseries 48} (2022) 18:1}.

\bibitem{alnaes2014unified}
M.S.~Aln{\ae}s, A.~Logg, K.B.~{\O}lgaard, M.E.~Rognes and G.N.~Wells,
  \emph{{Unified Form Language}: A domain-specific language for weak
  formulations of partial differential equations},
  \href{https://doi.org/10.1145/2566630}{\emph{ACM Transactions on Mathematical
  Software} {\bfseries 40} (2014) 9:1}.

\end{thebibliography}\endgroup


\newpage
\appendix
\section{Comments on Numerics}
In the main text, we have discussed the thermal and microcanonical OTOCs of a single quantum particle in various billiard systems. There were two numerical methodologies used in this work, corresponding to the work of section \ref{sec:qm_otoc} and section \ref{sec:b-field_otoc}, respectively.

\subsection{Numerics for Section \ref{sec:qm_otoc}}
Our language of implementation for  section \ref{sec:qm_otoc} was \textit{Mathematica}. In general, the challenge is to construct the $b_{nm}$ matrix, from which we can construct the microcanonical OTOC and the thermal OTOC. The successful construction of this matrix has three ingredients:
\begin{itemize}
    \item The energy spectrum, $\{E_n\}$;
    \item The position matrix, $x_{mn}$;
    \item The momentum matrix, $p_{mn}$.
\end{itemize}
As mentioned in the main text (see eq. \eqref{eq:bnm}), it is sufficient to have only the energy spectrum and the position matrix for constructing $b_{mn}$. For the Morse potential, one can use the formulae presented in the text (eq. \eqref{eq:morsespectrum1} and \eqref{eq:morseposition}) to construct $b_{mn}$ and compute the microcanonical OTOCs and thermal OTOCs therefrom, tuning the width $a$, depth $D_e$, and temperature $T$ to study the varying behaviour of the Morse system.

\subsection{Numerics for Section \ref{sec:b-field_otoc} \label{app:num3}}
\begin{figure}
    \centering
    \includegraphics[width=0.9\linewidth]{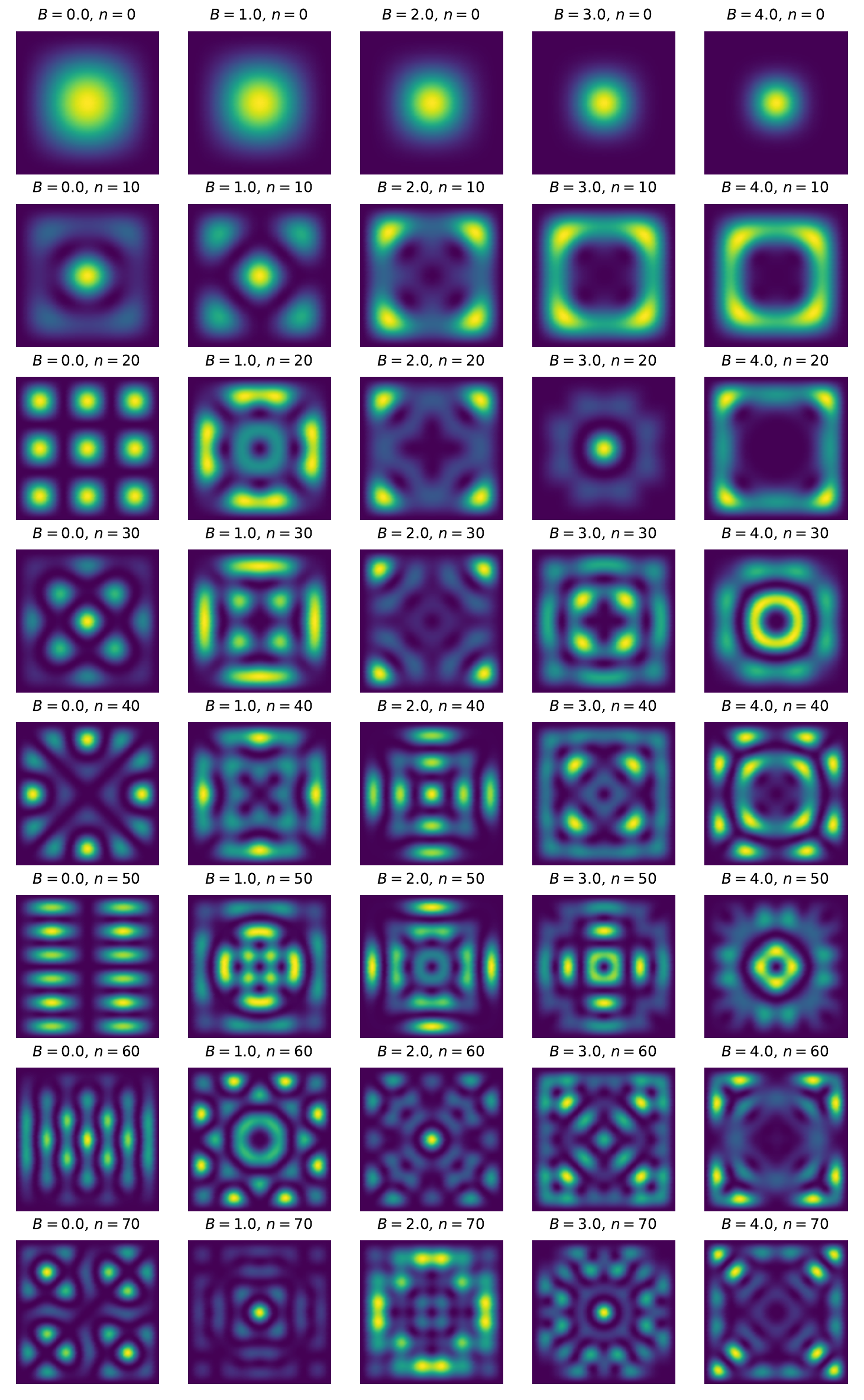}
    \caption{Probability density functions for a single particle in a square billiard for a range of transverse magnetic field strengths.}
    \label{fig:squarpd}
\end{figure}

\begin{figure}
    \centering
    \includegraphics[width=0.9\linewidth]{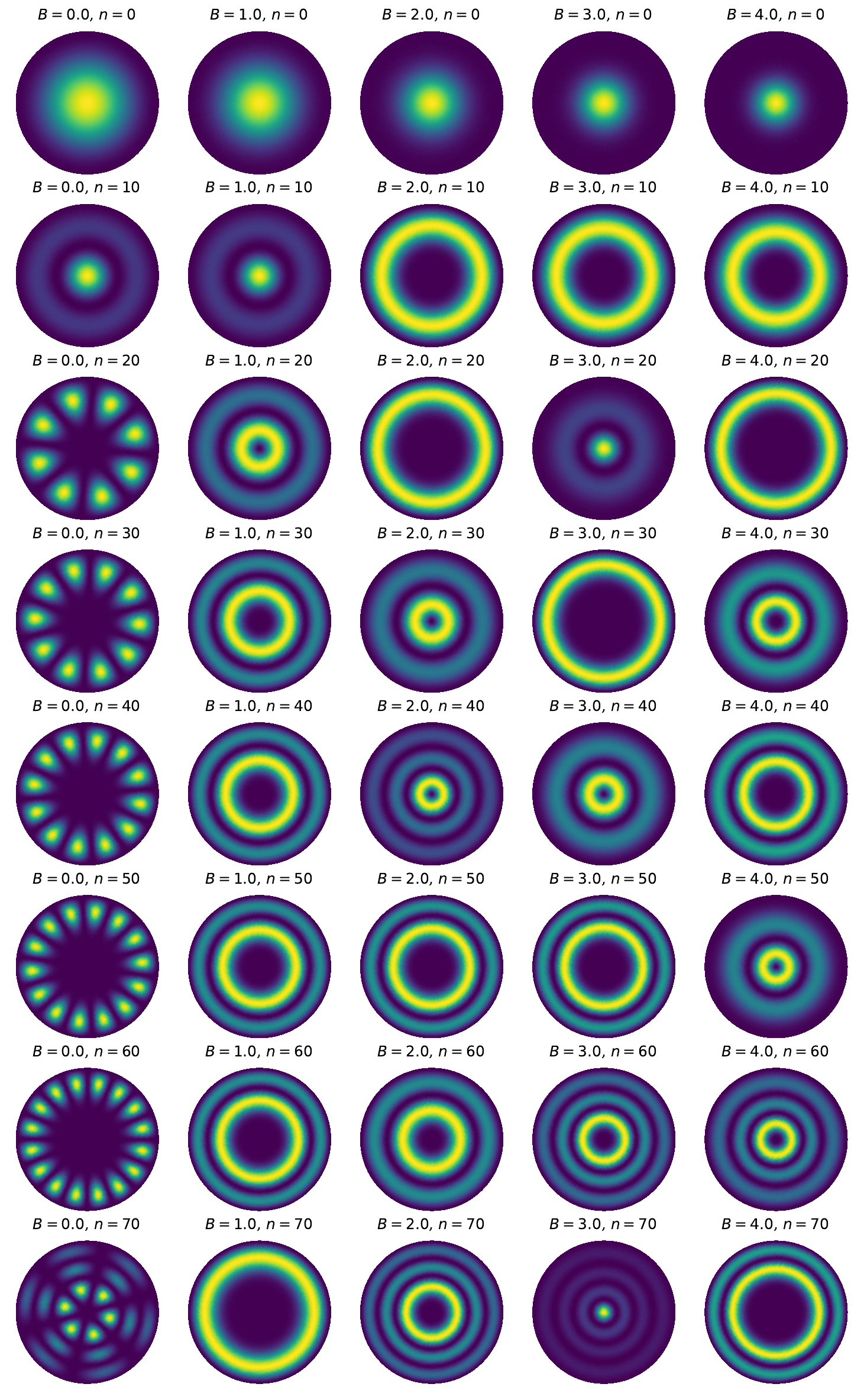}
    \caption{Probability density functions for a single particle in a circular billiard for a range of transverse magnetic field strengths.}
    \label{fig:diskpd}
\end{figure}

\begin{figure}
    \centering
    \includegraphics[width=0.9\linewidth]{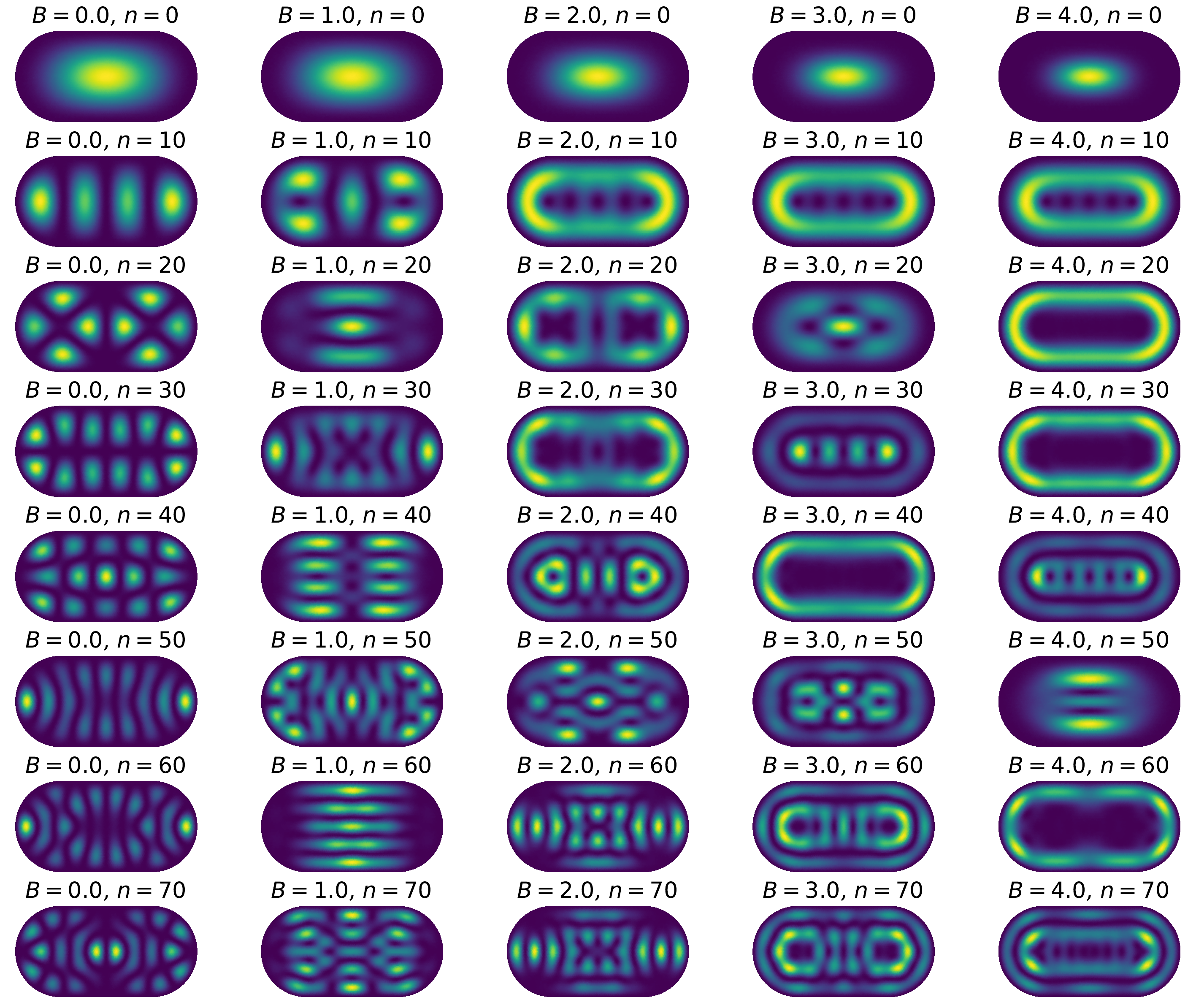}
    \caption{Probability density functions for a single particle in a stadium billiard for a range of transverse magnetic field strengths.}
    \label{fig:stadiumpd}
\end{figure}

Solving for the microcanonical OTOCs of general billiard configurations is a more involved process, requiring a finite-element method implemented in \textit{Python} and making extensive use of the \textit{FEniCS} library \cite{baratta2023dolfinx,scroggs2022basix,scroggs2022construction,alnaes2014unified}.
\\ \\
The process works as follows:
\begin{enumerate}
    \item One defines the geometry of the space (i.e. square/disk/stadium) and discretises the space, placing a mesh over the billiard. This creates a list of nodes (mesh vertices) and triangle (or other meshing shape) connectivity (assuming a triangular mesh, this stores the triplet of nodes that define a given triangle).
    \item A small `basis' function is placed at each node, $\phi_{node=i}(x,y)$. For our implementation, we use:\\   
    
    \verb|{basix.ufl.element("Lagrange", cell_type.cellname(), 2, shape=(2,))}|\\
    
    which in an instruction to be continuous across the boundaries of a mesh cell, to be a polynomial of maximum degree 2, and to be a 2-component vector (representing the real and imaginary components of the wavefunction). 
    \item The library assumes that the wavefunction can be approximated by $$\psi(x,y)=\sum_ic_i\phi_{node=i}(x,y),$$ and the solver's job is to solve for the $c_i$.
    \item You provide the \textit{weak form} of the Schrödinger equation; instead of requiring $H\ket{\psi}=E\ket{\psi}$, this equation is weaker, requiring that $\bra{v}H\ket{\psi}=E\braket{v|\psi}$, for any \textit{test} vector $\ket{v}$. For the integral representation of the Schrödinger equation, we write
    $$\int v(H\psi)\ dxdy=E\int v\psi\ dxdy.$$
    One uses integration by parts to convert terms as 
    $$ \int v \nabla^2\psi\rightarrow-\int(\nabla v)\cdot(\nabla\psi). $$
    This simplifies the continuity requirements at the boundary of a mesh cell.
    \item At every single mesh cell, the integral equation is generated. Since the basis functions are polynomials of maximum degree 2, this process is exceptionally fast. The outcome (for a triangular mesh) is a $(2\times6)\times(2\times6)$ matrix; the 2's being associated to the wavefunction having two components (real and imaginary) and 6's are contributed from the three basis functions at the nodes and the 3 midpoints of the edges between the nodes. This latter contribution is required to account for the curvature of a polynomial of degree 2, and these nodes are inserted by the algorithm as additional nodes, but do not have basis functions placed thereon. These contributions are collected from all triangles and summed into a global \textit{`Hamiltonian' matrix}, $A$ (the left-hand side of the integral equation, $\int v(H\psi)$), and a \textit{`mass' matrix}, $B$, (the integral component of the right-hand side of the integral equation, $\int v\psi$). For example, for one of the triangle cells in the mesh the contribution to $A$ is
    $$
    \begin{pmatrix}
        \int{\phi_{i_1}H\phi_{i_1}} & \cdots &\int{\phi_{i_1}H\phi_{i_6}}\\
        \vdots & & \vdots\\
        \int{\phi_{i_6}H\phi_{i_1}}&\cdots & \int{\phi_{i_6}H\phi_{i_6}}
    \end{pmatrix},
    $$
    where the contributions from the $\int{\phi_{i_n}H\phi_{i_m}}$ component are added to $A_{i_ni_m}$. The $B_{i_n i_m}$ contributions will be $\int \phi_{i_n}\phi_{i_m}$.
    \item The resultant global matrices, $A$ and $B$, are sparse since the majority of nodes will not be vertices of the same triangle cell. The library then attempts to solve the eigensystem
    $$
    A\mathbf{x} = \lambda B\mathbf{x}
    $$
    for the eigenvector, $\mathbf{x}$, and the eigenvalue, $\lambda$, using the $A$ and matrix and the \textit{mass matrix}, $B$. The $\phi_i$ are not true basis functions --  they overlap with their neighbours and are not orthogonal. Numerical eigensolvers are conventionally good at finding the \textit{largest} eigenvalues of a system. To instead solve for the \textit{smallest} eigenvalues, the system recasts this eigensystem with a shift-and-invert procedure,
    $$
    \underbrace{(A-\sigma B)^{-1}B }_M\ \mathbf{x}=\frac{1}{\lambda-\sigma}\mathbf{x},
    $$
    where $\sigma$ is a target eigenvalue to be near (in our implementation, $\sigma=0$).
    \item The algorithm defines a random vector, $\mathbf{v}$. It repeatedly applies the matrix $M$ to this vector, $M^n\mathbf{v}$, which has the effect of amplifying the largest eigenvalues of $M$, and eliminates non-eigenvalue components. In reality, the solver uses Krylov subspace projection to project the huge global matrices onto a Krylov subspace, and solves the eigensystem over this reduced space. 
    \item From this, the algorithm extracts $\lambda_i=E_i$ and $\mathbf{v}_i$. These eigenvectors correspond to the vector $\mathbf{c}_i$, present in the `basis' decomposition of the wavefunction \\$\psi_i(x,y)=\sum_n c_{in}\phi_{n}$, where the subscript on the wavefunction indicates the eigenstate for the $i^{th}$ eigenenergy.
\end{enumerate}
In this way, the solver can accurately numerically approximate the true eigenenergies and the corresponding eigenstates, from which we can construct the probability densities for a given billiard and a given energy. An important limitation of these solvers is that they often easily cluster around solutions at $E=0$ and $E=1$, producing erroneous solutions, which must be systematically removed. Resulting probability densities from using this method for the square, disk, and stadium billiard are shown in Figs. \ref{fig:squarpd}, \ref{fig:diskpd}, and \ref{fig:stadiumpd}, respectively.

\newpage

\end{document}